\documentclass[aps,pra,twocolumn,amsmath,amssymb,nofootinbib,showpacs,superscriptaddress, longbibliography]{revtex4-1}
\usepackage[english]{babel}
\usepackage{bbold}
\usepackage{latexsym}
\usepackage{graphics}
\usepackage{graphicx}
\usepackage{epsfig}
\usepackage{color}
\usepackage{bm}
\usepackage{amsmath}
\usepackage{amssymb}
\usepackage{amsthm}
\usepackage{dcolumn}
\usepackage{bm}
\usepackage{float}
\usepackage{hyperref}
\usepackage{color}
\usepackage{epstopdf}
\usepackage{braket}
\usepackage{cleveref}
\usepackage[svgnames]{xcolor}
\hypersetup{hidelinks,colorlinks=true,allcolors=DarkBlue}

\usepackage{comment}

\theoremstyle{remark}

\usepackage{algorithm}
\usepackage{algpseudocode}

\usepackage{tabularx, makecell, booktabs}

\usepackage{multirow}

\newcommand{\circuit}{\mathbf{c}}
\newcommand{\sample}{\mathbf{s}}

\newcommand{\bs}{\boldsymbol}

\usepackage[normalem]{ulem}
\usepackage{xcolor}
\usepackage[makeroom]{cancel}
\usepackage{todonotes}
\usepackage{tabularx}

\begin{document}

\preprint{APS/123-QED}
\date{\today}
\title{Continuous monitoring for noisy intermediate-scale quantum processors}
\author{Y.F. Zolotarev}
\affiliation{Russian Quantum Center, Skolkovo, Moscow 143026, Russia}
\affiliation{National University of Science and Technology ``MISIS”, Moscow 119049, Russia}
%\affiliation{Moscow Institute of Physics and Technology, Dolgoprudny, Moscow Region 141701, Russia}
\author{I.A.  Luchnikov}
\affiliation{Russian Quantum Center, Skolkovo, Moscow 143026, Russia}
\affiliation{National University of Science and Technology ``MISIS”, Moscow 119049, Russia}
\author{J.A. López-Saldívar}
\affiliation{Russian Quantum Center, Skolkovo, Moscow 143026, Russia}
\affiliation{National University of Science and Technology ``MISIS”, Moscow 119049, Russia}
%\affiliation{Moscow Institute of Physics and Technology, Dolgoprudny, Moscow Region 141701, Russia}
\author{Aleseky K. Fedorov}
\affiliation{Russian Quantum Center, Skolkovo, Moscow 143026, Russia}
\affiliation{National University of Science and Technology ``MISIS”, Moscow 119049, Russia}
\author{E.O. Kiktenko}
\affiliation{Russian Quantum Center, Skolkovo, Moscow 143026, Russia}
\affiliation{National University of Science and Technology ``MISIS”, Moscow 119049, Russia}
%\affiliation{Steklov Mathematical Institute of Russian Academy of Sciences, 8 Gubkina St., Moscow 119991, Russia}

\begin{abstract}
We present a continuous monitoring system for intermediate-scale quantum processors that allows extracting estimates of noisy native gate and read-out measurements based on the set of executed quantum circuits and resulting measurement outcomes. 
In contrast to standard approaches for calibration and benchmarking quantum processors, the executed circuits, which are input to the monitoring system, are assumed to be out of any control.
We provide the results of applying our system to the synthetically generated data obtained from a quantum emulator, as well as to the experimental data collected from a publicly accessible cloud-based quantum processor.
In the both cases, we demonstrate that the developed approach provides valuable results about inherent noises of emulators/processors.
Considering that our approach uses only already accessible data from implemented circuits without the need to run additional algorithms, the monitoring system can complement existing approaches. 
We expect that our monitoring system can become a useful tool for various quantum computers in the near-term horizon, including publicly accessible cloud-based platforms, 
and reduce resources that are required for their benchmarking and calibration. 
\end{abstract}

\maketitle

\section{Introduction}

Quantum computers are considered as a tool for providing speed up in solving various computational problems~\cite{Ladd2010,Aspuru-Guzik2021,Fedorov2022}, 
ranging from prime factorization~\cite{Shor1994} to simulating complex systems~\cite{Lloyd1996,Nori2014,Altman2021}. 
Currently, quantum devices are of intermediate size and significantly affected by noises, 
which manifests the era of noisy intermediate-scale quantum (NISQ) technology~\cite{Preskill2018,Fedorov2022,Aspuru-Guzik2021,Babbush2021-4}. 
Existing NISQ devices are actively used for searching quantum advantage~\cite{Martinis2021,Pan2020} and prototyping quantum algorithms~\cite{Babbush2021-4}, 
in particular, in the quantum chemistry domain~\cite{Aspuru-Guzik2019-2,Aspuru-Guzik2020}. 
Various quantum computing devices are accessible via cloud services~\cite{Soeparno2021}.
However, due to noise effects, the parameters of quantum computers, such as fidelity of state preparation, gates, and measurements, change with time significantly, which naturally affects the results of implementing quantum algorithms and limits the capabilities of quantum computing devices~\cite{Aspuru-Guzik2021,Fedorov2022}. 
This problem is typically tackled by calibrating quantum processors that involve standard techniques to characterize quantum states and processes, such as quantum tomography and special (randomized) benchmarking procedures~\cite{BlumeKohout2020,Gambetta2011,Gambetta2012,Zoller2022,nielsen2021gate}. 
Such procedures can be seen as ``service quantum algorithms'', which require either running additional quantum circuits (thereby they use the computational time of the devices) or sometimes stopping the workflow of quantum processors. 
With the increased amount of requests to quantum computing devices~\cite{Chong2022}, 
this problem will become more and more important. 

\begin{figure*} [ht!]
    \centering
    \includegraphics[width=0.9\linewidth]{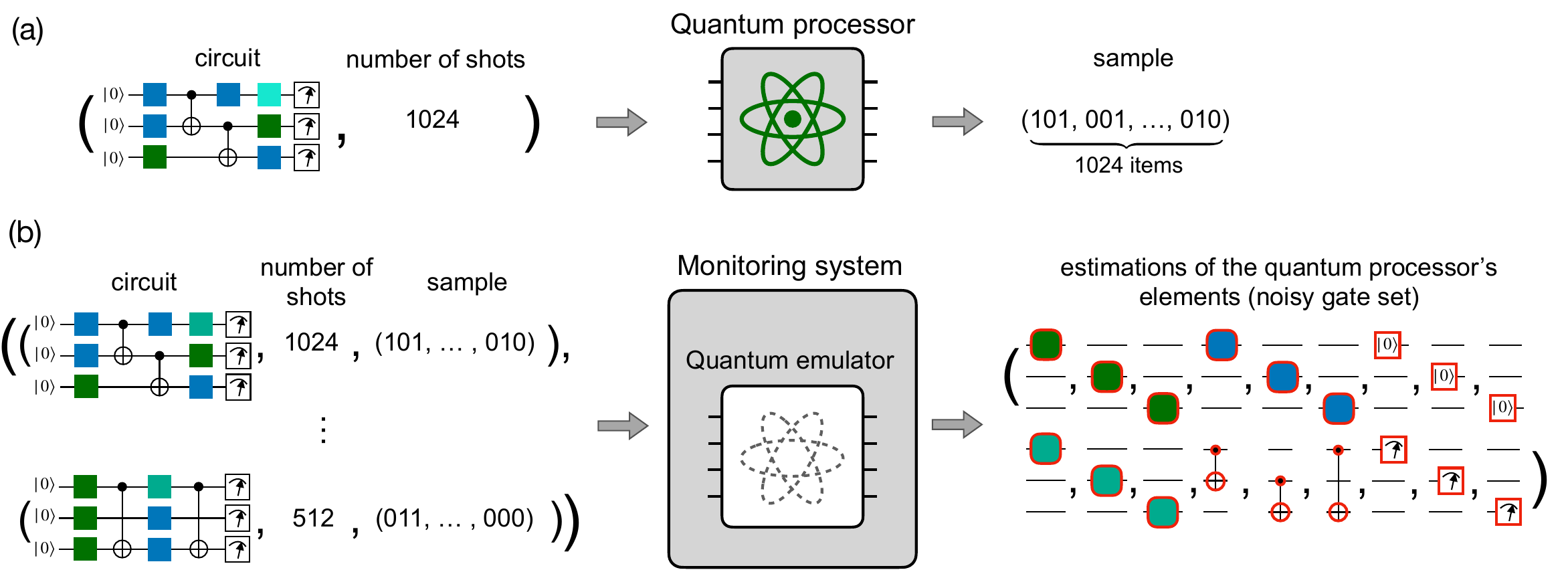}
    \caption{In (a) the concept of a quantum processor as black box is presented.
    The quantum processor takes a circuit ${\bf c}$, which consists of native gates, and number of times $K$ for the circuit to be run. As output, the processor yields $K$ bit strings obtained from read-out measurements.
    In (b) the general concept of the monitoring system is illustrated.
    The system takes a sequence of input-output pairs from the quantum processor and provides estimates on all basic units of the quantum processor. 
    Units include all native gates, state initialization, and read-out measurements. Estimates are obtained using a built-in quantum emulator.
    }
    \label{fig:general_scheme}
\end{figure*}

A complementary approach is to consider a quantum processor as a `black box' [see Fig.~\ref{fig:general_scheme}(a)] 
and use all the accessible information from the users of quantum computers for characterizing parameters of a quantum processor with certain accuracy. 
In this case, a set of quantum circuits that have been executed on a quantum processor and corresponding read-out results can be used. 
On the basis of these data, one can estimate the parameters of quantum computer components. 
A recent study~\cite{Chong2022} has analyzed the IBM quantum cloud over two years indicating that over 600,000 quantum circuit have been executed with 
almost 10 billion `shots' or trials over 20 quantum computational devices. 
It is then reasonable to expect that with the increase of the amount of implemented circuits, the accuracy of such reconstructions will be increased. 
An advantage of such an approach is that it uses only already accessible data from implemented circuits without the need of running additional algorithms that spend resources of quantum processors, e.g. the quantum gate set tomography~\cite{nielsen2021gate} or the randomized benchmarking~\cite{knill2008randomized}.
It is also important to note that within the `black box' approach the input circuits are assumed to be out of any control (e.g. they can be designed by remote users). 
However, the main challenge is to extract valuable information from the results of conducted quantum experiments.
Moreover, classical simulation/emulation of the workflow of quantum processors is required, which limits the size quantum processors that can be monitored using this framework.
One of the most advanced approach that can be efficiently used for this purpose is the tensor network approach~\cite{Orus2019-6}.

In the present work, we develop a continuous monitoring framework for intermediate-scale quantum processors that allows extracting estimates of noisy native gate and read-out measurements based on the set of executed quantum circuits and resulting measurement outcomes [see Fig.~\ref{fig:general_scheme}(b)]. 
The operation of the system is based on employing an inherent tensor-network-based emulator.
We provide technical details of the workflow of the monitoring system, and then show its performance on a synthetic and real data.
The synthetic data is generated using an additional emulator of a noisy 5-qubit processor. 
Since the form of all noisy gates is known in this case, we are able to compare to the results of the monitoring system with the corresponding `true values'.
We justify the correctness the monitoring system's operation by showing the increase of prediction accuracy with a growth of processed circuits number.
Then we apply the developed system to the data obtained from publicly accessible 5-qubit cloud-based superconducting processor.
In this case the true form noisy gates is unknown, yet we demonstrate the correctness and usefulness of the monitoring system by comparing its results with the data obtained from an additionally launch measurement calibration procedure.

Our work is organized as follows.
In Sec.~\ref{sec:concept}, we formalize an operation of a noisy quantum processor as a `black box'.
In Sec.~\ref{sec:design}, the implementation of the monitoring system is described.
In Sec.~\ref{sec:application}, we apply the developed system to synthetically generated and real data.
The conclude and discuss avenues of a future research and development in Sec.~\ref{sec:application}

\section{Noisy quantum processors operation}\label{sec:concept}

One can think about the quantum processor as a `black box' taking certain inputs and providing outputs on these inputs, as it is shown in Fig.~\ref{fig:general_scheme}(a).
An input to the quantum processor can be represented in the form of a quantum circuit $\circuit$ that is paired together with number of times (shots) $K$ it has to be executed.
Circuit $\circuit$ can be considered as a list of (elementary) operations performed on $n$ processor's qubits, labeled by indices $0,\ldots,n-1$, and formally can be represented as an ordered set 
\begin{equation}
    {\bf c} = ((l_0,t_0),\ldots,(l_{{\rm g} - 1},t_{\#{\rm g} - 1})),
\end{equation}
where $l_i$ denotes the label of the particular operation,
$t_i$ denotes target qubit(s) at which the operation acts, and $\#{\rm g}$ is the total number of elementary operations in the circuit.
In what follows, we consider single-qubit and two-qubit elementary operations, so targets $t_i$ can be represented either as a qubit's index ($t_i\in\{0,\ldots, n-1\}$), or as a pair of two distinct indices ($t_i=(t_i^1, t_i^2)$, $t_i^j\in\{0,\ldots, n-1\}$), correspondingly.

We consider two types of operations: (i) single-qubit state preparation and measurement (SPAM) operations and (ii) realization of native gates.
We assume that all qubits are initially prepared in the state $\ket{0}$ that corresponds to the first $n$ entries in $\circuit$ to be taken in the form $({\sf P}, i)$ for $i=0,\ldots,n-1$.
The last $1\leq m\leq n$ entries of $\circuit$ determine single-qubit computational basis measurements and have a special label $l_i={\sf M}$ (the corresponding target $t_i\in\{0,\ldots,n-1\}$ defines a qubit to be measured).
The set of native gates is assumed to be consisted of the finite number of single-qubit and two-qubit gates (i.e. $l_i\in \{{\sf H},{\sf S}, {\sf T}, {\sf CX}\}$), which is a common practice for existing quantum processors. 
We note that ${\bf c}$ is the classical object, so it can be  stored in the memory of a classical computer efficiently.

The corresponding output is yielded as a list of $K$ bit strings ${\bf s}=(s^1,\ldots,s^K)$, where the length of each bit string $s^i$ is given by the number of qubits $m$ measured in the end of the circuit.
We then refer to $\sample$ as a sample produced by circuit $\circuit$.
The essence of the quantum processor's operation is that output bit strings are generated from a certain probability distribution.
In the ideal (noiseless) case, this distribution is determined by the sequence of unitary operators specified by gates in the input circuit.
However, due to various imperfections causing decoherence, the actual probability distribution for output bit strings in real NISQ devices appears to be different from the ideal expected one.

Within our work on the development of the monitoring system, we impose the following assumptions about the noisy behaviour of the real quantum processor and processed circuits.
First, we assume that noisy behaviour of elementary operations in a real processor can be described by assigning each operation with a completely positive trace-preserving (CPTP) map (quantum channel). 
In this way, we suppose that errors coming from realizations of each individual operation are independent.
Second, though CPTP maps, corresponding to same gate acting on different qubits, can be different (e.g., a noisy Hadamard gate acting on qubit 1 may differ from a noisy Hadamard gate acting on qubit 2), 
we assume that the error fallibility of the gate does not depend on its time position in the circuit (i.e., a noisy Hadamard gate on qubit 1 in the beginning of the circuit is the same as the noisy Hadamard gate on qubit 1 in in the end of the circuit).

Taking into account these assumptions, we can write down an expression for the probability to obtain sample $\sample$ from running circuit $\circuit$.
It is convenient to introduce a concept of a gate set.
By a gate set ${\bf G}$, we understand a set of triples of the following form:
\begin{equation}
    {\bf G} = \{(l,t,\Phi_l^t)\},
\end{equation}
where $l$ and $t$ are the gate label and the target of elementary operation, respectively (exactly as in the case of circuits), and $\Phi_l^t$ is a CPTP map of the corresponding operation.
In the case of the native gates realization, $\Phi_l^t$ describes the implementation of the gate with label $l$ on target $t$. 
In the case of SPAM operations, $\Phi_{\sf P}^t$ is a CPTP map acting on the ideal initial state $\ket{0}$, and $\Phi_{\sf M}^t$ is also a CPTP map preceding the ideal protective single-qubit computational basis measurement of qubit $t$.
From the practical point of view, it is convenient to represent $\Phi_l^t$ in the form of a Choi matrix~\cite{jamiolkowski1972linear, choi1975completely, jiang2013channel}, 
i.e., $\Phi_l^t$ of a single-qubit (two-qubit) gate can be defined by $4\times 4$ ($16\times 16$) matrix satisfying standard conditions on Choi matrices.
We then refer to the channel $\Phi_l^t[\cdot]$ that corresponds to the gate with label $l$ and target $t$, taken from the gate set ${\bf G}$ as ${\bf G}(l,t)[\cdot]$.
By $\widetilde{{\bf G}}(l,t)[\cdot]$ we define a CPTP map that acts in the whole space of $n$ qubits: 
\begin{equation}
    \widetilde{{\bf G}}(l,t)[\cdot] = \left({\bf G}(l,t)_t \otimes {\rm Id}_{\overline{t}}\right) [\cdot],
\end{equation}
where ${\bf G}(l,t)_t$ denotes ${\bf G}(l,t)$ acting in the space of target qubit(s) $t$, and ${\rm Id}_{\overline{t}}$ denotes an identity map acting in the space of other $n-1$ (or $n-2$) qubits.
Using the introduced notations, the whole CPTP map ${\cal F}_{\circuit}({\bf G})[\cdot]$ that is realized by circuit ${\bf c}$ with gates from ${\bf G}$ is given by
\begin{equation}
    {\cal F}_{\bf c}({\bf G})[\cdot]=
    \widetilde{\bf G}(l_{\#{\rm g}- 1},t_{\#{\rm g} - 1})
    \circ 
    \ldots  \\
    \circ \widetilde{\bf G}(l_{0},t_{0})[\cdot],
\end{equation}
where $\circ$ stands for the standard composition.

The probability of obtaining a sample ${\bf s}=(s^1,\ldots,s^K)$ after $K$ shots of circuit $\circuit$ realized with a quantum processor, whose operations are described by gate set ${\bf G}$, is given by
\begin{equation} \label{eq:prob}
    P({\bf s}|{\bf c}, {\bf G})= \prod_{j=1}^K P(s^j|{\bf c}, {\bf G}),
\end{equation}
where
\begin{equation} \label{eq:prob-outcome}
     P(s^j|{\bf c}, {\bf G})= {\rm Tr}\left(\ket{s^j}\bra{s^j}
        {\cal F}_{\circuit}({\bf G})[\rho_{\rm in}]]\right)
\end{equation}
is th probability to obtain a bit string $s^j$, $\rho_{\rm in}=(\ket{0}\bra{0})^{\otimes n}$ is the fixed ideal initial state, and $\ket{s^j}\bra{s^j}$ is the shorthand notation of the projector on $\ket{s^j}$ state in the space of measured qubits, i.e. qubits with ${\sf M}$ gate in the end of the circuit ${\bf c}$ (in the remaining space, $\ket{s^j}\bra{s^j}$ acts as an identity).
We note that logarithm of $P({\bf s}|{\bf c}, {\bf G})$ can be written in a compact form as follows:
\begin{multline}  \label{eq:logprob}
     \log P({\bf s}|{\bf c}, {\bf G})= \sum_{j=1}^K
        \log{\rm Tr}\left(\ket{s^j}\bra{s^j}
        {\cal F}_{\circuit}({\bf G})[\rho_{\rm in}]]\right)\\
        =\sum_{s\in {\bf s}} \#s \log{\rm Tr}\left(\ket{s}\bra{s}
        {\cal F}_{\circuit}({\bf G})[\rho_{\rm in}]]\right),
\end{multline}
where $\#s$ is the number of times a bit string $s$ appears in sample ${\bf s}$.

\section{Design of the monitoring system}\label{sec:design}

Consider a quantum processor operating in a continuous regime, i.e. sequentially providing output samples $\sample_1, \sample_2, \ldots$ by processing a stream of input circuits with corresponding shots numbers $(\circuit_1, K_1), (\circuit_2, K_2), \ldots$.
Importantly, here we assume that input circuits $\circuit_i$ themselves are beyond the control: 
The particular circuits executed on the processor can be formed by remote users, while the monitoring system is launched by an administrator of the service, providing access to the quantum processor.

The goal of the monitoring system is in the gathering the actual information about the quantum processor without borrowing its computation time on running specific protocols.
Consider a time moment right after the obtaining measurement outcomes for the $i$th circuit.
At this moment the input to the monitoring system is given by the following expression:
\begin{equation} \label{eq:D_i}
    {\bf D}_{i}^M = ({\bf d}_{\max(1,i-M)},\ldots, {\bf d}_i), 
    \quad     
    {\bf d}_j =  ({\bf c}_j, K_j, {\bf s}_j),
\end{equation}
where $M$ is the hyperparameter determining the maximal number of circuits to be processed by the monitoring system.
One can set $M=\infty$ that corresponds to the processing of all previously launched circuits.

We note that due to the fact that the noise level of real quantum processor typically changes in time, it is a common practice to periodically suspend the operation of the processor for a maintenance and calibration purposes.
In the case of such maintenance procedure, the introduced circuit counter is reset to zero.

The output of the monitoring system to the input ${\bf D}_{i}^M$ is gate set ${\bf G}^{\rm est}_{i}$ that contain estimates of the `true' gate set of the quantum processor.
The estimated gate set is obtained as a solution of the following optimization problem
\begin{equation}~\label{eq:opt_problem}
    {\bf G}^{\rm est}_i = \arg\min_{{\bf G}} {\cal L}({\bf G}|{\bf D}_i^M),
\end{equation}
where the minimization is performed over CPTP maps inside a gate set ${\bf G}$, and the cost function is given by
\begin{multline} \label{eq:cost_func}
    {\cal L}({\bf G}|{\bf D}_i^M)= \\ 
    -\sum_{j=\max(1,i-M)}^{i} \log P({\bs s}_{j}|\circuit_{j},{\bf G})
    +{\cal D}^{\rm reg}_{\lambda_1,\lambda_2}({\bf G}).
\end{multline}
Here, the first term is the standard log-likelihood function of the gate set ${\bf G}$ with respect to the input data ${\bf D}_i^M$. 
The second term in Eq.~\eqref{eq:cost_func} serves for regularization purposes and reads
\begin{multline} \label{eq:reg}
    {\cal D}^{\rm reg}_{\lambda_1,\lambda_2}({\bf G})=\sum_{q=1,2}\lambda_q \sum_{l,t: q\text{-qubit}} \|{\cal C}({\bf G}(l,t))-\\{\cal C}({\bf G}^{\rm ideal}(l,t))\|_{\rm F}^2,
\end{multline}
where $\lambda_1,\lambda_2\geq 0$ are regularization parameters, $\|\cdot\|_{\rm F}$ stands for the Frobenius norm, ${\cal C}(\Phi)$ is the Choi matrix of channel $\Phi$, and ${\bf G}^{\rm ideal}$ is the gate set consisted of the flawless realization of all operations. 
The Choi matrix for a single-qubit channel is defined as 
\begin{equation}
    {\cal C}(\Phi) = \sum_{j,j'=0,1} \Phi[\ket{j}\bra{j'}] \otimes \ket{j}\bra{j'},
\end{equation}
and for two-qubit one as
\begin{equation}
    {\cal C}(\Phi) = \sum_{j,j',k,k'=0,1} \Phi[\ket{jk}\bra{j'k'}] \otimes \ket{jk}\bra{j'k'}.
\end{equation}
All CPTP maps within ${\bf G}^{\rm ideal}$ are represented by an action unitary operators
\begin{equation}
    {\bf G}^{\rm ideal}(l,t)[\rho] = U_l\rho U_l^\dagger,
\end{equation}
where $U_l$ is a unitary operator of the gate with label $l$.
The flawless SPAM operation corresponds to $U_{\sf P}=U_{\sf M}= \mathbb{1}$, where $\mathbb{1}$ is the identity matrix.

The essence of the regularization term~\eqref{eq:reg} is the distance between the elements of the estimated gate set ${\bf G}^{\rm est}_i$ and the ideal gate set ${\bf G}^{\rm ideal}$.
This term has a probabilistic interpretation of the contribution of the gate set prior probability distribution concentrated near the ideal gate set.
Indeed, one expects the true gate set to be close to the ideal gate set.
Note that in Eq.~\eqref{eq:reg} we go separately through all single-qubit and two qubit-gates with different regularization parameters $\lambda_1$ and $\lambda_2$ correspondingly.

One of the main challenges related to the solving Eq.~\eqref{eq:opt_problem} is to compute the cost function ${\cal L}({\bf G}|{\bf D}_i^M)$ that requires obtaining probabilities for given outcomes in the form~\eqref{eq:prob-outcome}.
This is a point where the inherent quantum emulator comes into play.
We develop our own tensor networks-based quantum emulator that allows one to compute exact form of Eq.~\eqref{eq:logprob} for circuits of low width or depth.
The operation of the emulator is based on contraction of a tensor-network specified by the circuit, gate set elements, and the particular measurement outcome.
We note that no approximations, i.e. the limitation of tensor ranks, is used while contraction.

The need in the emulator poses the main limitation of the developed system; specifically, the monitoring system can process classically simulatable circuits only.
However, this is the case for a variety of existing developing quantum processors. 
Moreover, there are various techniques allowing simplifications in the simulation of quantum processors in the quantum advantage regime~\cite{Martinis2019} (we place this discussion in Sec.~\ref{sec:concl}). 
As we demonstrate in the next section, the developed monitoring system copes with a task of monitoring existing five-qubit superconducting processors.
We also provide additional benchmarks of the developed tensor-network-based emulator in Appendix~\ref{sec:app:benchs}.

The minimization procedure in Eq.~\eqref{eq:opt_problem} with respect to the elements of ${\bf G}$, namely quantum channels $\{\Phi_l^t\}$, is realized by the gradient based Riemannian optimization technique~\cite{boumal2020introduction, absil2009optimization} 
that preserves CPTP constraints on the elements of ${\bf G}$. 
To calculate gradient of the loss function defined in Eq.~\eqref{eq:cost_func}, we utilize automatic differentiation~\cite{liao2019differentiable}. 
In particular, the Riemannian adaptive moment estimation (RAdam) algorithm~\cite{becigneul2018riemannian, li2020efficient} is efficient.
The implementation of the RAdam algorithm for CPTP constraints is taken from \textsc{QGOpt} package~\cite{luchnikov2021qgopt}.

\section{Application of the monitoring system} \label{sec:application}

Here we consider an application of the developed monitoring system to the cloud-based 5-qubit quantum processor.
We perform two kinds of numerical experiments: `synthetic' and `real' ones.

In the synthetic experiment, we replace the quantum processor under observation with a noisy quantum emulator with known noise parameters and known noisy gate set (do not confuse this emulator with the emulator inside the monitoring system).
The goal here is to make sure that the monitoring system is able to learn the noise parameters of the emulator from the results of running random circuits, which is impossible to strictly confirm without prior knowledge of these parameters.
In the real experiment the input data for the monitoring system comes from the publicly-accessible superconducting processor \textsc{ibmq\_lima}.

To achieve the consistency between two experiments, they take place under the same conditions. 
(i) The emulator in the synthetic experiment is considered to have the same coupling topology and native gate set as in the real processor.
(ii) Synthetic input circuits and their corresponding shots numbers to produce the output samples are the same in the both experiments.

The rest of the section is organized as follows.
First, we describe the architecture of the quantum processor under consideration and its noisy emulator.
Next, we describe the construction of input circuits.
Finally, we consider the results of synthetic and real numerical experiments.

\subsection{Architecture of the processor}\label{sec:processor}

\begin{figure} [!ht]
    \centering
    \includegraphics[width=0.4\linewidth]{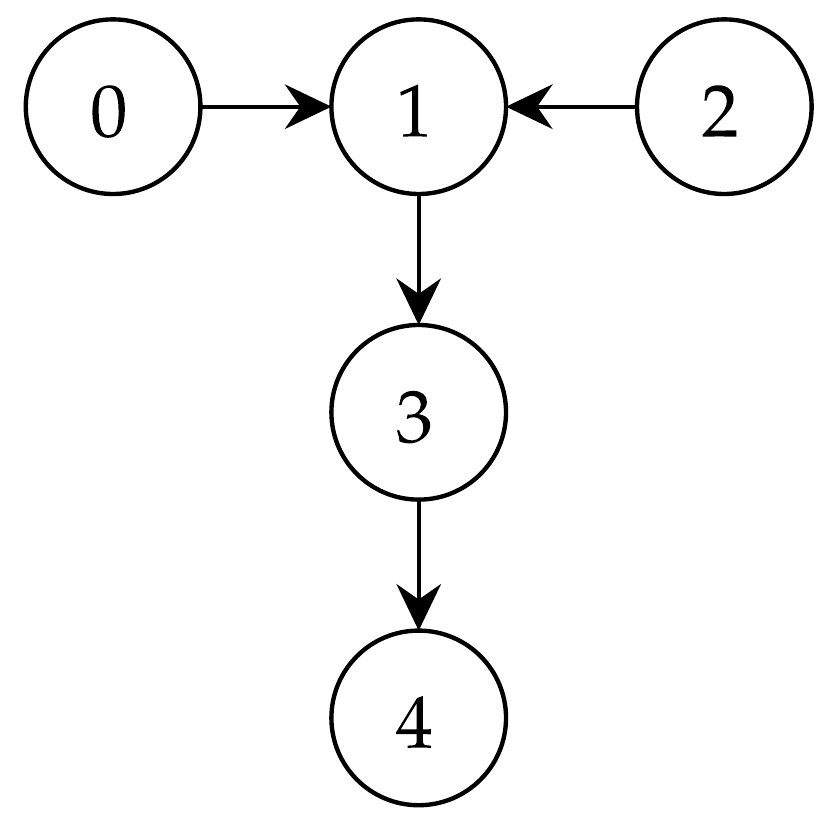}
    \caption{Architecture of the considered 5-qubit processor \textsc{ibmq\_lima}.
    The arrows corresponds to an ability to perform ${\sf CX}$ gate: Arrows are directed from control qubits to target ones.}
    \label{fig:processor}
\end{figure}

We consider a 5-qubit quantum processor \textsc{ibmq\_lima} with a coupling map shown in Fig.~\ref{fig:processor}.
Each arrow in Fig.~\ref{fig:processor} corresponds to the ability to perform a native two-qubit ${\sf CX}$ (controlled NOT) gate
\begin{equation}
    U_{\sf CX} = 
    \begin{bmatrix}
        1&0&0&0\\0&1&0&0\\0&0&0&1\\0&0&1&0
    \end{bmatrix}.
\end{equation}
Though the both directions of ${\sf CX}$ gates within connection in the coupling map are allowed by the architecture of the processor, we fix the ${\sf CX}$ directions as it in shown in Fig.~\ref{fig:processor}: 
Arrows are directed from control qubits to target ones.

As native single-qubit gates, which can be applied to every qubit, we consider a finite set
\begin{equation}
    \begin{aligned}
        & U_{\sf RZ} = \begin{bmatrix}
            e^{-\imath\pi/8} & 0\\ 0 & e^{\imath\pi/8}
        \end{bmatrix},
        & U_{\sf X} = \begin{bmatrix}0&1\\1&0\end{bmatrix},\\
        & U_{\sf SX}=\frac{1}{2} \begin{bmatrix}
            1 + \imath & 1 - \imath \\
            1 - \imath & 1 + \imath
        \end{bmatrix}, 
        & U_{\sf ID}=\begin{bmatrix}1&0\\0&1\end{bmatrix}.
    \end{aligned}
\end{equation}
We note that the introduced gate set is universal and can be used for constructing any desired 5-qubit unitary operation up to preset accuracy.

\subsection{Noisy emulator}\label{sec:emulator}

In order to construct the noisy emulator of the considered processor, we assign to each native gate acting on a qubit (or a pair of qubits) its own quantum channel.
In this way, we obtain a known true noisy gate set, which we denote as ${\bf G}^{\rm true}$.

The construction of quantum channels is motivated by standard approach used for modelling noisy behavior of near-term quantum computers~\cite{georgopoulos2021modeling}.
In particular, we take into account a longitudinal ($T_1$) and transverse ($T_2$) relaxation of qubit's states, fluctuations of a length and/or intensity of control pulses, and general qubit errors modelled by depolarizing channel.

In the case of a single-qubit gate $U_l$, the corresponding quantum channel is obtained in the form:
\begin{multline} \label{eq:single-qubit_dec}
    {\bf G}^{\rm true}(l,t)
    [\cdot]= \Phi^{\rm ap}(\gamma^l_t,\mu^l_t)\circ \Phi^{\rm dep}(p^l_t)\circ \\
    \Phi^{\rm sm}(\nu^l_t;U_l) \circ \Phi^{\rm ap}(\gamma^l_t,\mu^l_t)[\cdot].
\end{multline}
Here
\begin{equation}
    \Phi^{\rm ap}(\gamma,\mu)[\cdot]=\Phi^{\rm a}(\mu)\circ\Phi^{\rm p}(\gamma)[\cdot]    
\end{equation}
is a sequence of a phase damping 
\begin{multline}
    \Phi^{\rm p}(\mu)[\rho] = \begin{bmatrix}
        1 & 0\\0&\sqrt{1-\mu} 
    \end{bmatrix}
    \rho
    \begin{bmatrix}
        1 & 0\\0&\sqrt{1-\mu} 
    \end{bmatrix}+\\
    \begin{bmatrix}
        0 & 0\\0&\sqrt{\mu} 
    \end{bmatrix}
    \rho
    \begin{bmatrix}
        0 & 0\\0&\sqrt{\mu} 
    \end{bmatrix}
\end{multline}
and an amplitude damping
\begin{multline}
    \Phi^{\rm a}(\gamma)[\rho] = \begin{bmatrix}
        1 & 0\\0&\sqrt{1-\gamma} 
    \end{bmatrix}
    \rho
    \begin{bmatrix}
        1 & 0\\0&\sqrt{1-\gamma} 
    \end{bmatrix}+\\
    \begin{bmatrix}
        0 & \sqrt{\gamma}\\0&0 
    \end{bmatrix}
    \rho
    \begin{bmatrix}
        0 & 0 \\ \sqrt{\gamma}&0 
    \end{bmatrix}
\end{multline}
channels, where $\gamma, \mu\in[0,1]$ are decoherence parameters; 
$\Phi^{\rm dep}(p)[\cdot]$ is a standard depolarization channel
\begin{equation}
    \Phi^{\rm dep}(p)[\rho] = (1-p)\rho + \frac{p}{2} {\rm Tr}[\rho] \mathbb{1}
\end{equation}
with depolarization parameter $p\in[0,1]$;
and 
\begin{equation} \label{eq:smooth1}
    \begin{split}
        & \Phi^{\rm sm}(\nu;U)[\rho] = \int e^{-\imath H(U) \tau} \rho e^{\imath H(U) \tau} \omega_\nu(\tau) {\rm d}\tau
    \end{split}
\end{equation}
with
\begin{equation} \label{eq:smooth2}
    \omega_\nu(\tau)=\frac{1}{\sqrt{2\pi\nu}}\exp{\frac{-(\tau-1)^2}{2\nu^2}}, \quad
        H(U) = \imath\log(U),
\end{equation}
corresponds to implementing $U$ by acting with the effective Hamiltonian $H(U)$ over time period fluctuating according to Gaussian distribution with unit mean and standard deviation $\nu$.
Note that $\Phi^{\rm sm}(\nu;U)$ does not affect identity channel with $U=\mathbb{1}$.
The noise model for ${\sf M}$ gate is also obtained by Eq.~\eqref{eq:single-qubit_dec}.
We note that the imperfections of SPAM operations are considered to be described by measurement ${\sf M}$ gates only (the preparation is considered to be ideal). 
We use the same assumption for the data from the real processor.

The channels for realization of two-qubit gates are obtained in similar way:
\begin{multline}
    {\bf G}^{\rm true}({\sf CX},t)[\cdot]=
     \Phi^{\rm ap}(\gamma^{\sf CX}_t,\mu^l_t)^{\otimes 2}
     \circ \Phi^{\rm dep}(p^{\sf CX}_t)^{\otimes 2}\circ \\
    \Phi^{\rm sm}(\nu^{\sf CX}_t;U_{\sf CX}) \circ \Phi^{\rm ap}(\gamma^{\sf CX}_t,\mu^{\sf CX}_t)^{\otimes 2}[\cdot].
\end{multline}
Here the two-qubit target $t=(t^1,t^2)$ is taken in agreement with the coupling map shown in Fig.~\ref{fig:processor}, amplitude damping and depolarizing channels act on both qubits, and a `smoothed' ${\sf CX}$ realization $\Phi^{\rm sm}(\nu^{\sf CX}_t;U_{\sf CX})$ is given by Eq.~\eqref{eq:smooth1}, \eqref{eq:smooth2}.

\begin{table*} [htp]
    \centering
    \begin{tabular}{c|c|c|c|c|c|c}
         $l$ & $t$ & $\nu_t^l$ & $p^l_t$ &  $\mu^l_t$ & $\gamma^l_t$ & $F^l_t$  \\ \hline \hline
         ${\sf ID}$ & 0 & 0 & 0.015 & 0.01 & 0.015 & 0.980 \\
         ${\sf ID}$ & 1 & 0 & 0.01 & 0.022 & 0.012 & 0.979 \\
         ${\sf ID}$ & 2 & 0 & 0.003 & 0.013 & 0.01 & 0.989 \\
         ${\sf ID}$ & 3 & 0 & 0.005 & 0.005 & 0.009 & 0.992 \\
         ${\sf ID}$ & 4 & 0 & 0.01 & 0.006 & 0.005 & 0.988 \\ \hline
         ${\sf RZ}$ & 0 & 0.01 & 0.012 & 0.01 & 0.015 & 0.982 \\
         ${\sf RZ}$ & 1 & 0.16 & 0.03 & 0.022 & 0.012 & 0.960 \\
         ${\sf RZ}$ & 2 & 0.03 & 0.007 & 0.013 & 0.010 & 0.986 \\
         ${\sf RZ}$ & 3 & 0.02 & 0.01 & 0.005 & 0.009 & 0.988 \\
         ${\sf RZ}$ & 4 & 0.024 & 0.015 & 0.006 & 0.005 & 0.984 \\
    \end{tabular}
    \begin{tabular}{c|c|c|c|c|c|c}
         $l$ & $t$ & $\nu_t^l$ & $p^l_t$ &  $\mu^l_t$ & $\gamma^l_t$ & $F^l_t$  \\ \hline \hline
         ${\sf X}$ & 0 & 0.012 & 0.012 & 0.01 & 0.015 & 0.982 \\
         ${\sf X}$ & 1 & 0.015 & 0.008 & 0.022 & 0.012 & 0.980 \\
         ${\sf X}$ & 2 & 0.11 & 0.033 & 0.013 & 0.01 & 0.939 \\
         ${\sf X}$ & 3 & 0.014 & 0.007 & 0.005 & 0.009 & 0.990 \\
         ${\sf X}$ & 4 & 0.025 & 0.005 & 0.006 & 0.005 & 0.991 \\\hline
         ${\sf M}$ & 0 & 0 & 0.022 & 0.04 & 0.05 & 0.952 \\
         ${\sf M}$ & 1 & 0 & 0.013 & 0.012 & 0.015 & 0.981 \\
         ${\sf M}$ & 2 & 0 & 0.048 & 0.03 & 0.035 & 0.942 \\
         ${\sf M}$ & 3 & 0 & 0.007 & 0.012 & 0.016 & 0.985 \\
         ${\sf M}$ & 4 & 0 & 0.05 & 0.045 & 0.042 & 0.931 \\
    \end{tabular}
    \begin{tabular}{c|c|c|c|c|c|c}
         $l$ & $t$ & $\nu_t^l$ & $p^l_t$ &  $\mu^l_t$ & $\gamma^l_t$ & $F^l_t$  \\ \hline \hline
         ${\sf SX}$ & 0 & 0.18 & 0.017 & 0.01 & 0.015 & 0.960 \\
         ${\sf SX}$ & 1 & 0.02 & 0.01 & 0.022 & 0.012 & 0.978 \\
         ${\sf SX}$ & 2 & 0.01 & 0.005 & 0.013 & 0.01 & 0.987 \\
         ${\sf SX}$ & 3 & 0.01 & 0.007 & 0.005 & 0.009 & 0.990 \\
         ${\sf SX}$ & 4 & 0.013 & 0.013 & 0.006 & 0.005 & 0.986 \\ \hline
         ${\sf CX}$ & (0,1) & 0.05 & 0.005 & 0.006 & 0.008 & 0.978 \\
         ${\sf CX}$ & (1,3) & 0.05 & 0.01 & 0.005 & 0.009 & 0.971 \\
         ${\sf CX}$ & (2,1) & 0.01 & 0.01 & 0.007 & 0.013 & 0.972 \\
         ${\sf CX}$ & (3,4) & 0.06 & 0.03 & 0.02 & 0.03 & 0.917 \\
         \multicolumn{1}{l}{~}
    \end{tabular}
    \caption{Noise parameters of the emulator and resulting fidelities for all the considered native gates.}
    \label{tab:noise_pars}
\end{table*}

The chosen noise parameters are presented in Tab.~\ref{tab:noise_pars}.
We note that in order to make the emulator to be more realistic, we set the strengths of amplitude damping $\gamma_t^{l}$ and phase  damping $\mu_t^{l}$ to be the same for gates acting on the same qubit $t$.
In Tab.~\ref{tab:noise_pars} we also provide the resulting fidelities $F^l_t$ of noisy gates with respect to their ideal implementations given by
\begin{equation}
    \begin{split}
        & F^l_t = \frac{1}{4} \bra{\Psi_l}
        {\cal C}[{\bf G}^{\rm true}(l,t)]
        \ket{\Psi_l},\\
        & \ket{\Psi_l}=\sum_{j}U_l\otimes \mathbb{1}\ket{j,j}
    \end{split}
\end{equation}
in the single-qubit case, and by
\begin{equation}
    \begin{split}
        & F^l_t = \frac{1}{16} \bra{\Psi_l}
        {\cal C}[{\bf G}^{\rm true}(l,t)]
        \ket{\Psi_l},\\
        & \ket{\Psi_l}=\sum_{j,k}U_l\otimes \mathbb{1}^{\otimes 2}\ket{j,k,j,k}
    \end{split}
\end{equation}
in the two-qubit case.

For efficient sampling of output bit strings the technique described in Appendix~\ref{sec:app:emulator} is used.

\subsection{Input circuits}\label{sec:circuits}

\begin{figure} [t]
    \centering
    \includegraphics[width=\linewidth]{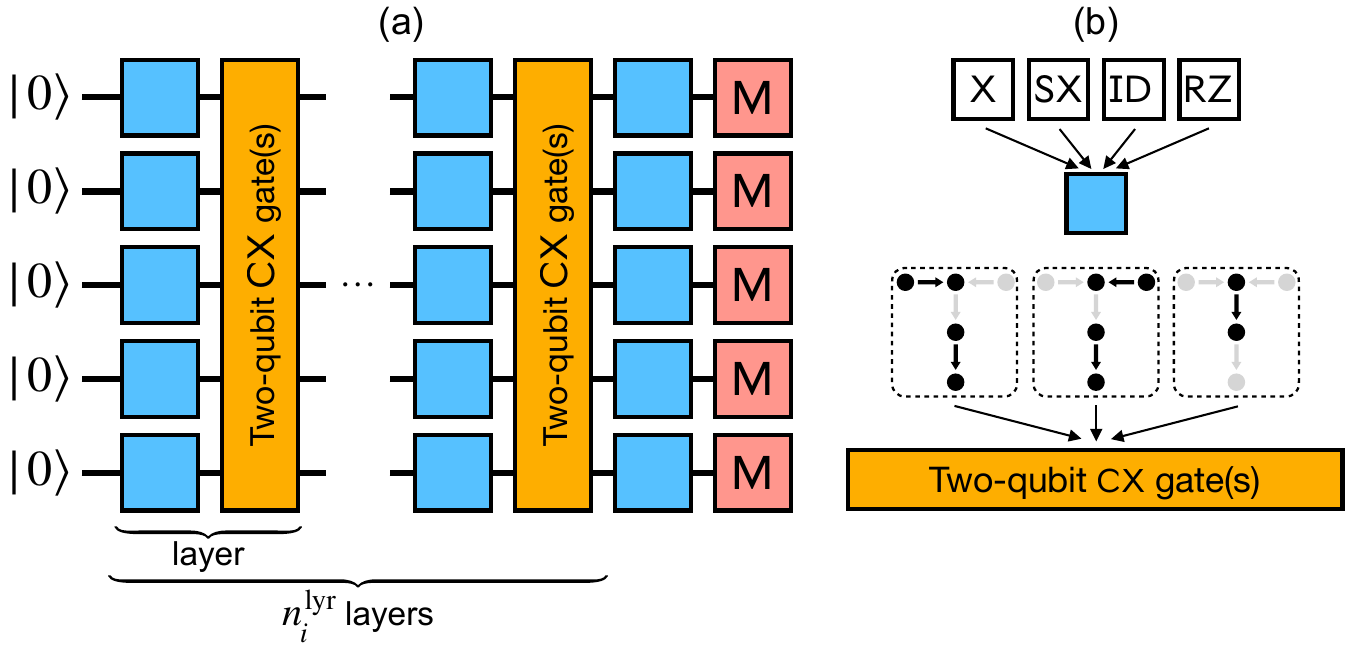}
    \caption{In (a) the structure of random circuits used for generating input data for the monitoring system is shown.
    In (b) the variants for a random choice for single-qubit gates and patterns of {\sf CX} gates in the two-qubit gates sublayers are presented.}
    \label{fig:random_circuits}
\end{figure}

We test the monitoring system on a set of random circuits generated in the form presented in Fig.~\ref{fig:random_circuits}(a).
Each circuit $\circuit_i$ has a layer structure, where each layer has two sublayers consisting of single-qubit and two-qubit gates correspondingly.
In the end of each circuit we put an additional sublayer of single-qubit gates followed by measurements of all qubits.
The number of layers $n^{\rm lyr}_i$ for circuit $\circuit_i$ is taken uniformly an random from the set $\{0,1,\ldots,10\}$.
Single qubit gates are taken uniformly at random from the set $\{{\sf X},{\sf SX}, {\sf RZ}, {\sf ID}\}$.
The pattern of ${\sf CX}$ gates implemented in two-qubit gates sublayers is also taken uniformly at random from set of three possible patterns shown in the bottom of Fig.~\ref{fig:random_circuits}(b).

\subsection{Monitoring the noisy emulator}\label{sec:monitoring-emulator} 

First, we apply the developed monitoring system to the data generated with the described emulator.
We generate $1024$ random circuits $\circuit_i$ (according to Sec.~\ref{sec:circuits}) and for each circuit get $K_i=8192$ bit strings of length 5 packed into a sample ${\bf s}_i$.
Then we apply the monitoring system to the input datasets ${\bf D}^\infty_i$ for $i=2^1,2^2,\ldots,2^{10}$ and obtain the corresponding estimated gates sets ${\bf G}^{\rm est}_i$.
We set regularization parameters $\lambda_1=100$, $\lambda_2=200$.
We refer to the circuits $\circuit_1,\ldots,\circuit_i$ as a training set, and to all other circuits as a test set.

To characterize the ability of the monitoring system to predict an output probability distribution, we consider a quantity
\begin{equation} \label{eq:l1_dist}
    \Delta^{\rm circ}_i(j) = \sum_s |P(s|\circuit_j,{\bf G}^{\rm est}_i) - f(s|{\bf s}_j)|,
\end{equation}
where the summation is performed over all possible bit strings $s\in\{0,1\}^5$, and $f(s|{\bf s}_j)=\#s_j/K_j$ is a frequency of appearing of $s$ in the resulting sample ${\bf s}_j$.
The essence of $\Delta^{\rm circ}_i(j)$ is an L1 distance between the observed distribution $f_j(s)$ obtained for the $j$-th circuit and the predicted distribution $P(s|\circuit_j,{\bf G}^{\rm est}_i)$ obtained after learning the monitoring system on $i$ circuits.

\begin{figure} [!ht]
    \centering
    \includegraphics[width=\linewidth]{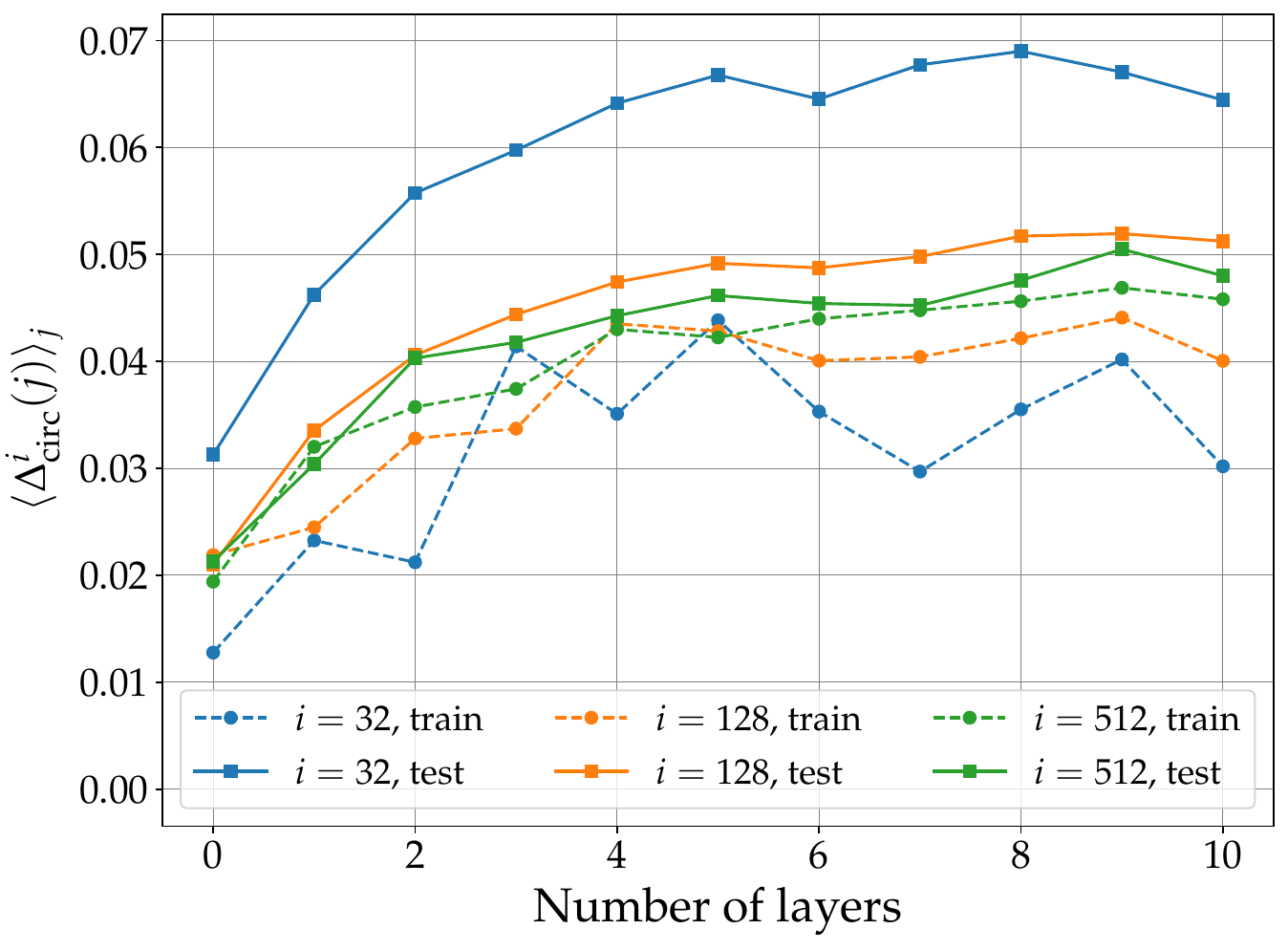}
    \caption{Mean values of prediction inaccuracy~\eqref{eq:l1_dist} obtained by averaging over circuits with the same number of layers inside the training and test sets for different number of training circuits $i$ (1024 circuits are considered in total).
    The input data for the monitoring system is generated with the described noisy quantum emulator.
    }
    \label{fig:l1norm_synth}
\end{figure}

The averaged values of $\Delta^{\rm circ}_i(j)$ are shown in Fig.~\ref{fig:l1norm_synth}.
Here the averaging is performed over circuits with the same layer number $n^{\rm lyr}_j$ separately for circuits inside the train set ($j\leq i)$ and inside the test set ($j>i$).
One can see that the accuracy of prediction increases with the growths of the training set, yet the difference between the length of the training set $i=128$ and $i=512$ is quite small.
Moreover, we see that the training curve and the test curve approach each other with the growth of $i$, and are almost equal for $i=512$.
This result suggest to pick up the value of the hyper parameter $M$ (the maximal number of processed circuits) on the level of 128-256.

We then consider how the monitoring system reconstructs individual gates.
We introduce two quantities, namely, 
\begin{equation}
    \begin{aligned}
        &\Delta_\diamond^{\rm true}(l,t;i) =
        \| {\bf G}^{\rm est}_i(l,t) - {\bf G}^{\rm true}(l,t)
        \|_\diamond,\\
        &\Delta_\diamond^{\rm ideal}(l,t;i) =
        \| {\bf G}^{\rm est}_i(l,t) - {\bf G}^{\rm ideal}(l,t)
        \|_\diamond,   \\
    \end{aligned}
\end{equation}
(where $\|\cdot\|_\diamond$ is the diamond norm) that provide a distance between an estimate gate ${\bf G}^{\rm est}_i(l,t)$ obtained after training on $i$ circuits and the corresponding true and ideal versions.
Remind that the distance between the real and estimated gates can be obtained only for the synthetically generated data.
We also note the monitoring system is not aware about the particular decoherence model chosen for $\{{\bf G}^{\rm true}(l,t)\}$ and performs reconstruction of $\{{\bf G}^{\rm est}_i(l,t)\}$ in the form of general CPTP maps.

For the special case of ${\sf M}$ gate, describing an imperfection of a read-out measurement, we consider quantities
\begin{equation}
    \begin{aligned}
        &\Delta_{\rm POVM}^{\rm true}(t;i) = \|M_0({\bf G}^{\rm est}_i({\sf M},t))-M_0({\bf G}^{\rm true}({\sf M},t)) \|_1,\\
        &\Delta_{\rm POVM}^{\rm ideal}(t;i) = \|M_0({\bf G}^{\rm est}_i({\sf M},t))-M_0({\bf G}^{\rm ideal}({\sf M},t)) \|_1, \\
    \end{aligned}
\end{equation}
where $\|\cdot\|_1$ stands for the standard L1 norm, and 
\begin{equation}
    M_k(\Phi)=
    {\rm Tr}_2\left[(\ket{k}\bra{k} \otimes \mathbb{1}) {\cal C}[\Phi]\right]
\end{equation}
is a POVM effect resulting from placing single-qubit channel $\Phi$ before the single-qubit computational basis measurement and corresponding to the outcome $k$ (here ${\rm Tr}_2$ denotes a partial trace with respect to the second qubit).
The values of $\Delta_{\rm POVM}^{\rm ideal}(t;i)$ ($\Delta_{\rm POVM}^{\rm true}(t;i)$) describe the maximal possible difference between probability distributions appearing in estimated and ideal (estimated and real) measurements.

\begin{figure*} [!ht]
    \centering
    \includegraphics[width=0.32\linewidth]{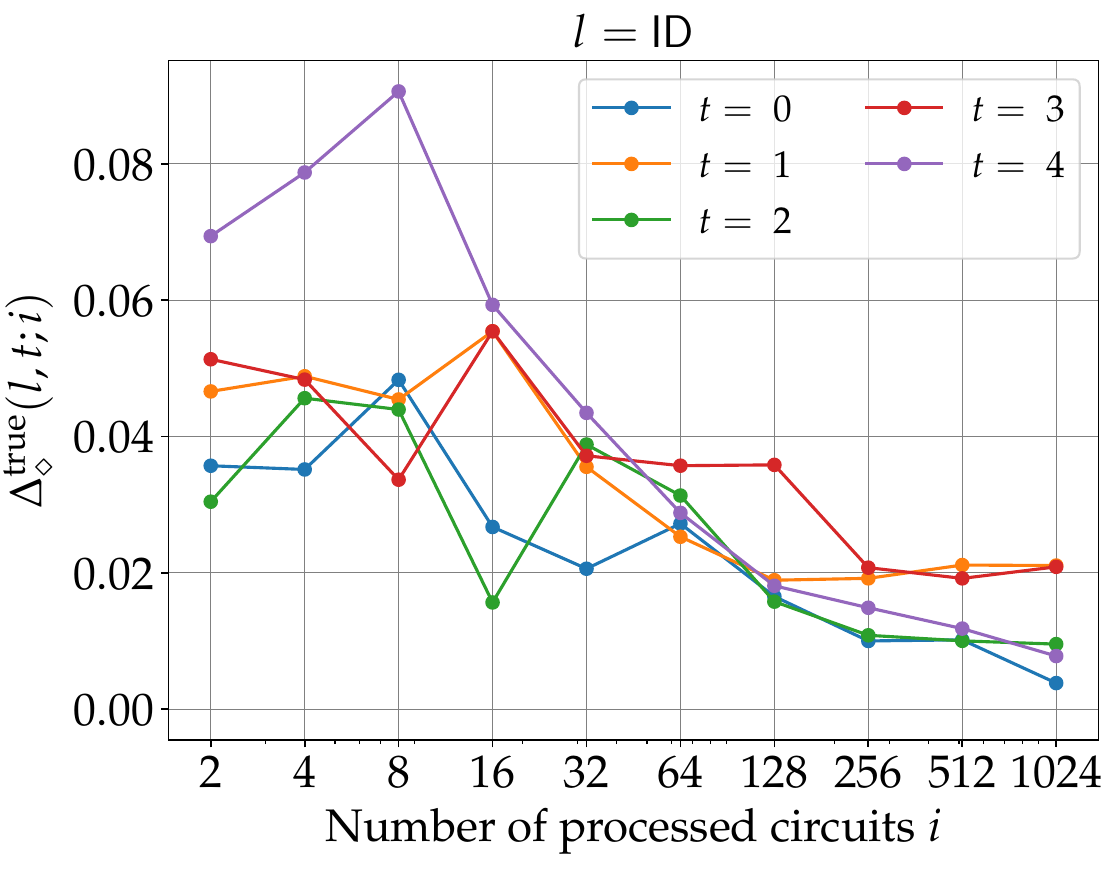}
    \includegraphics[width=0.32\linewidth]{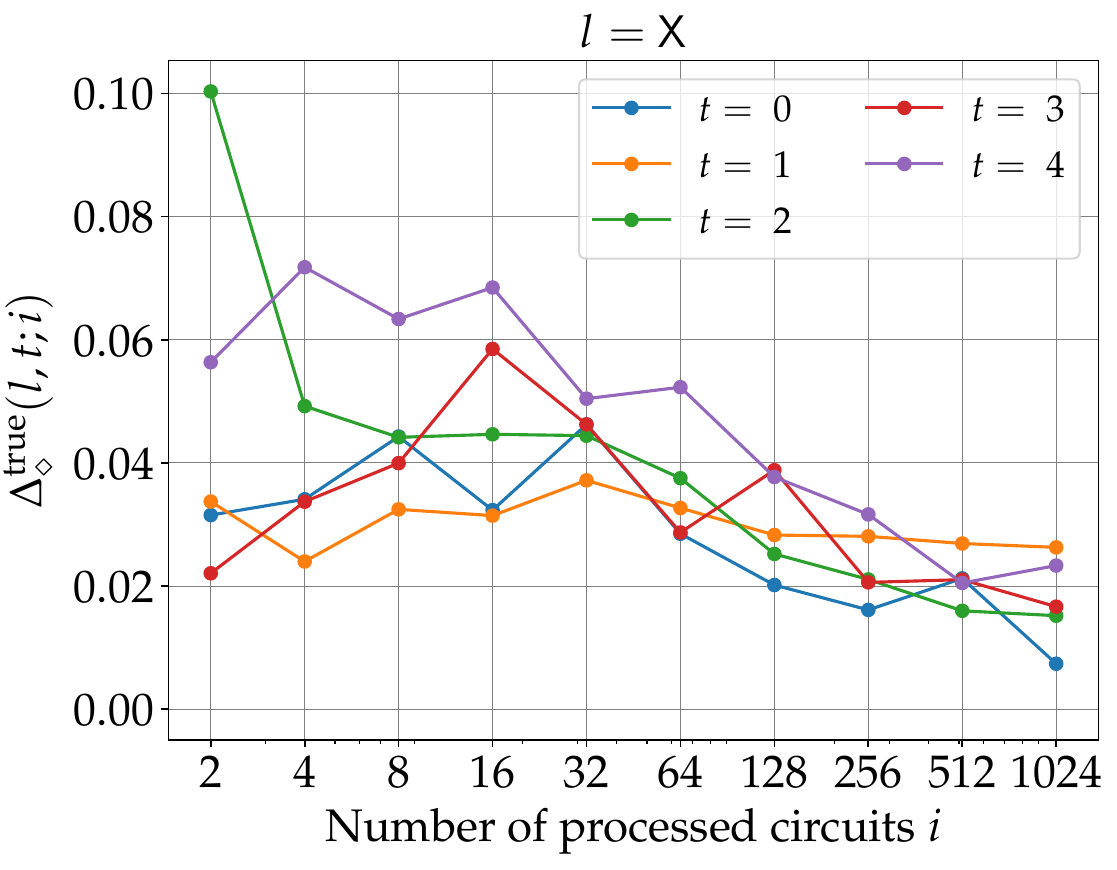}
    \includegraphics[width=0.32\linewidth]{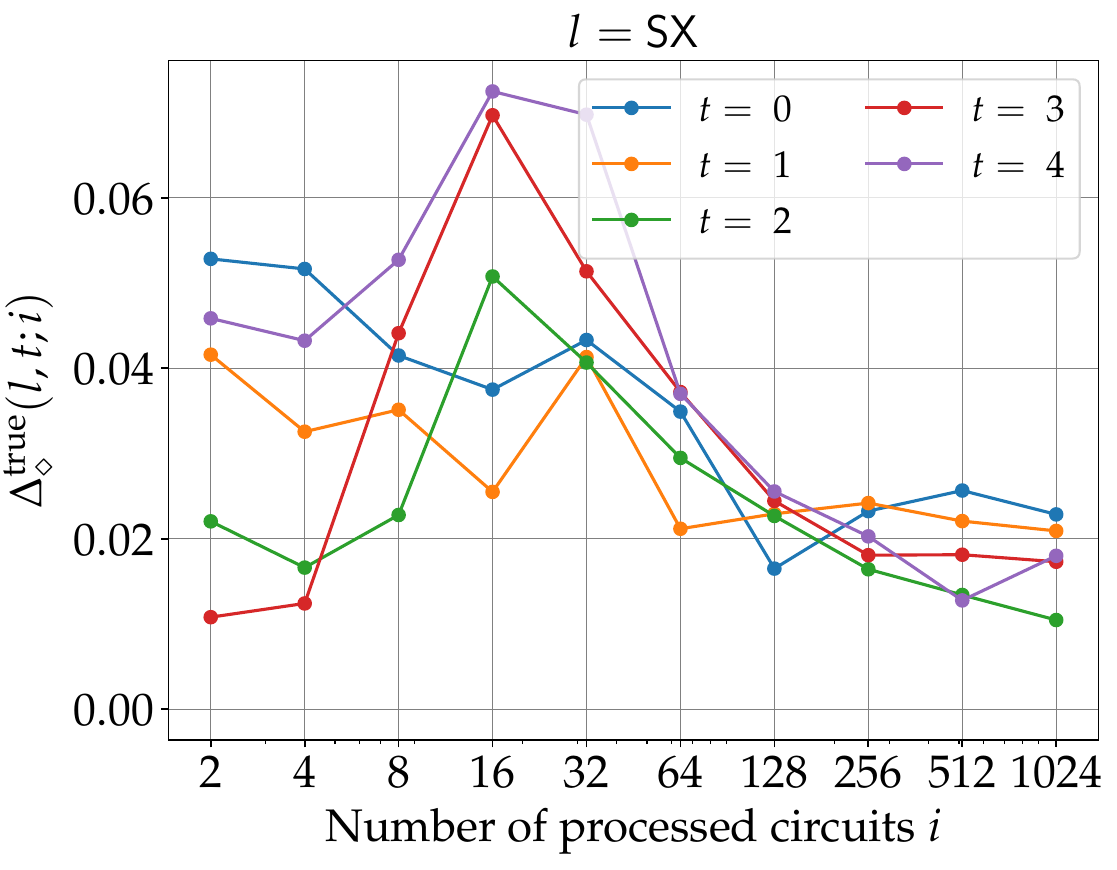}
    \includegraphics[width=0.32\linewidth]{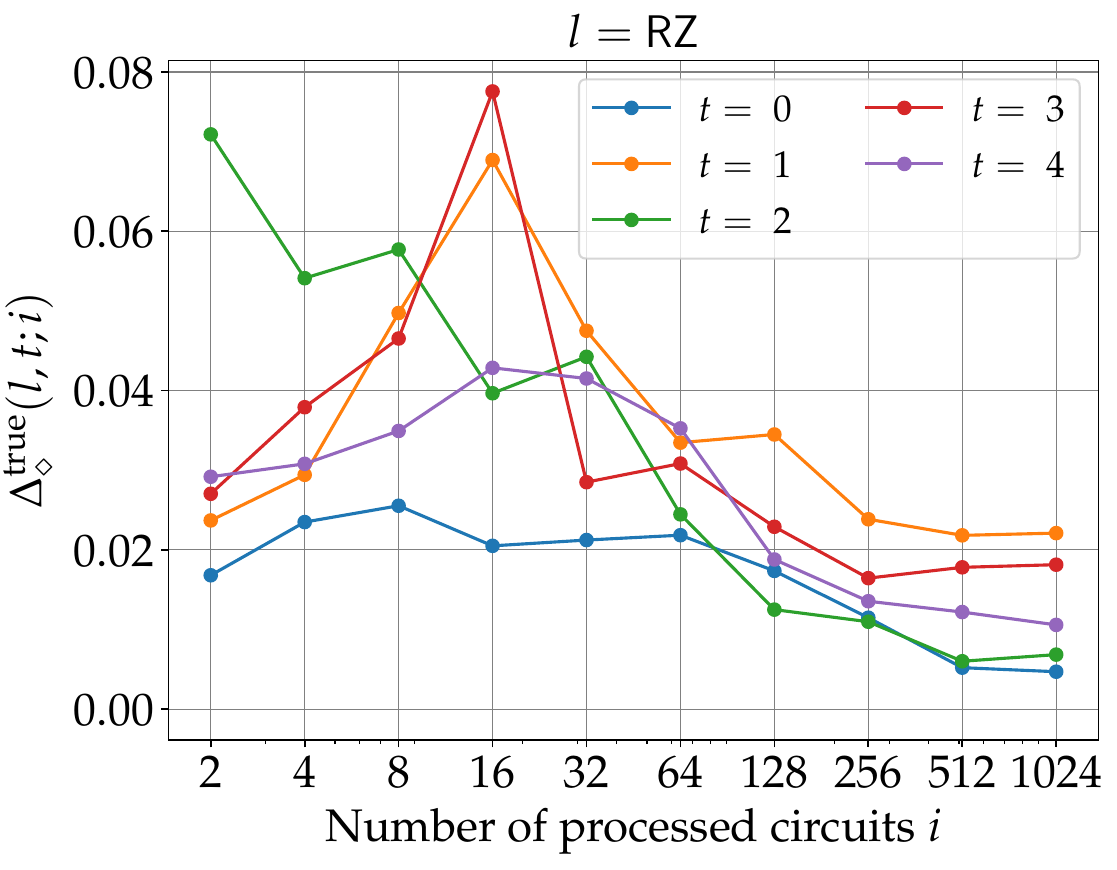}
    \includegraphics[width=0.32\linewidth]{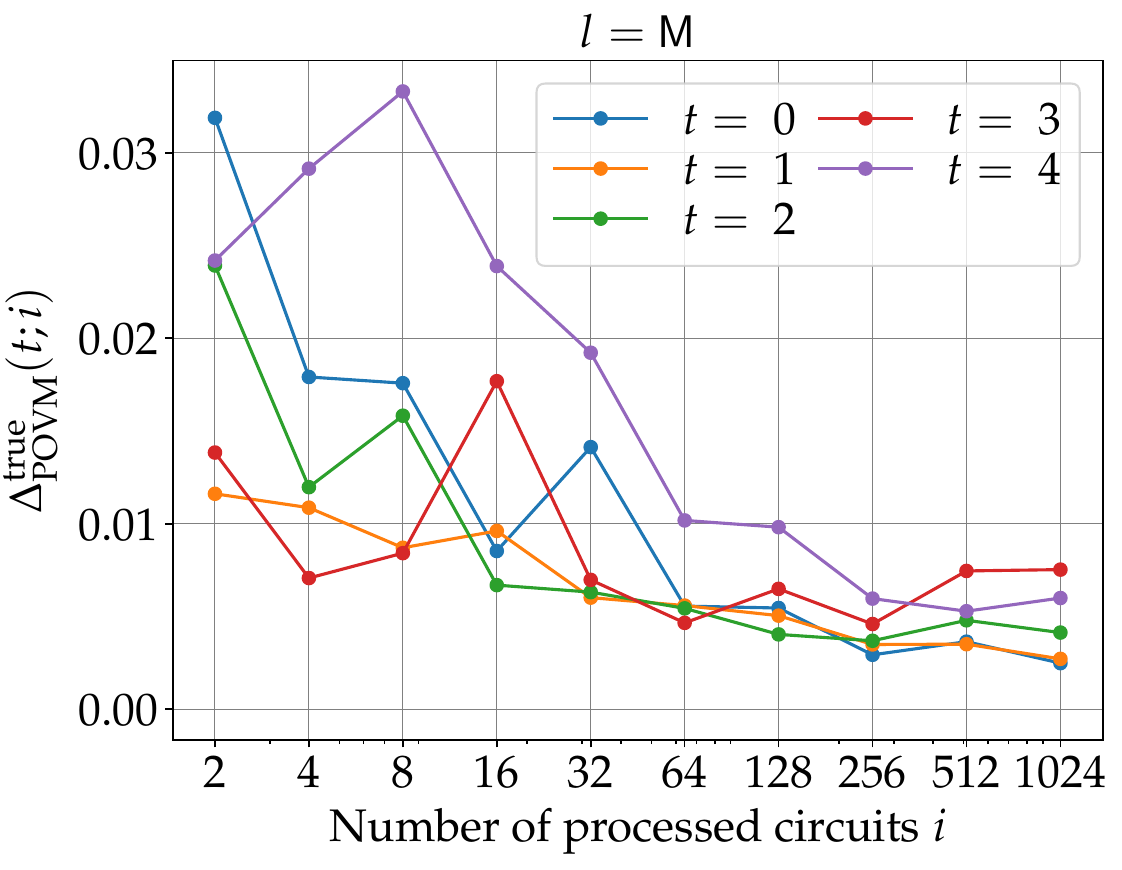}
    \includegraphics[width=0.32\linewidth]{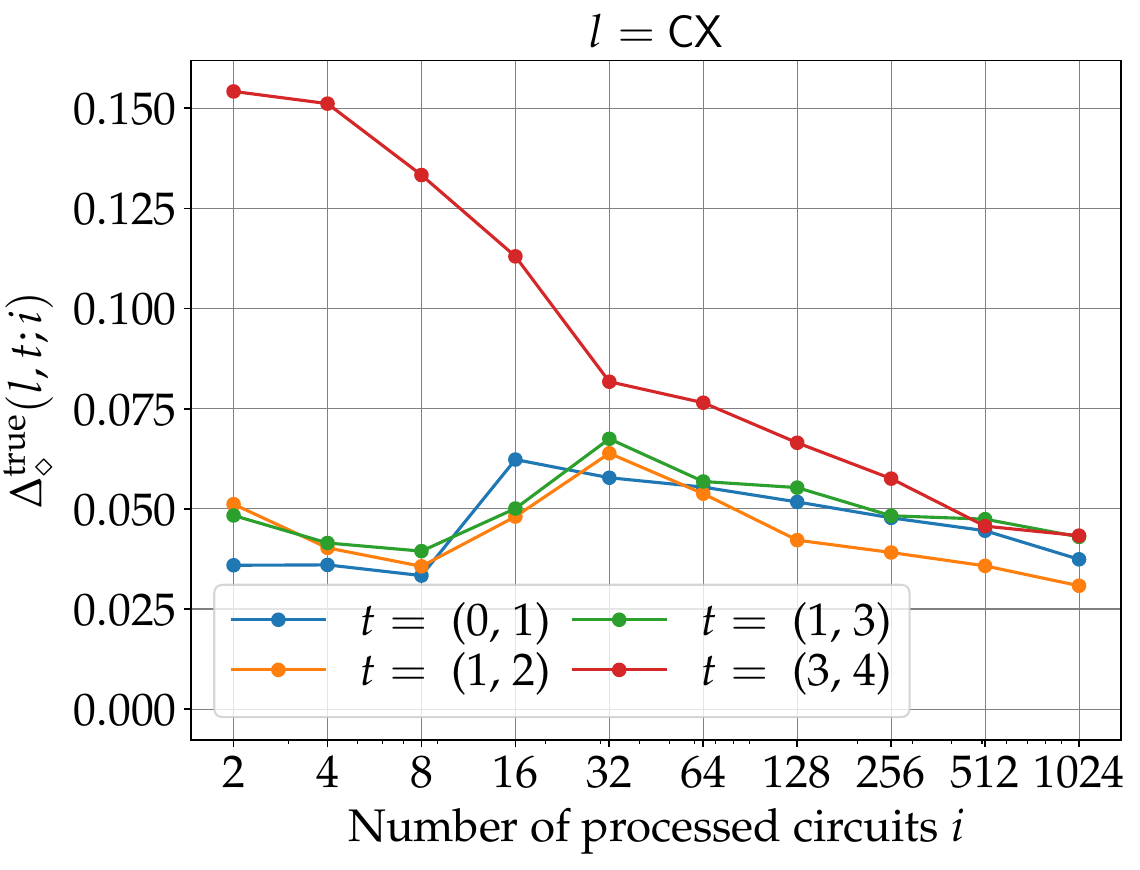}
    \caption{
    The distances between the true and estimated realizations of single-qubit gates (${\sf ID}$, ${\sf X}$, ${\sf SX}$, ${\sf RZ}$), two-qubit gates $({\sf CX}$), and read-out measurements (${\sf M}$) for different (pairs of) qubits 
    are shown as function of the number of processed circuits $i$.
    The input data for the monitoring system is generated with the described noisy quantum emulator.}
    \label{fig:synth_est_VS_real}
\end{figure*}

The distances between the elements of the estimated gate sets ${\bf G}^{\rm est}_i$ and the true gate set ${\bf G}^{\rm true}$ as functions of processed circuit number $i$ are presented in Fig.~\ref{fig:synth_est_VS_real}.
One can see that that all the curves tend to the zero value with a growth of processed cicuits $i$, that justifies the proper operation of the monitoring system. 

\begin{figure*} [!ht]
    \centering
    \includegraphics[width=0.32\linewidth]{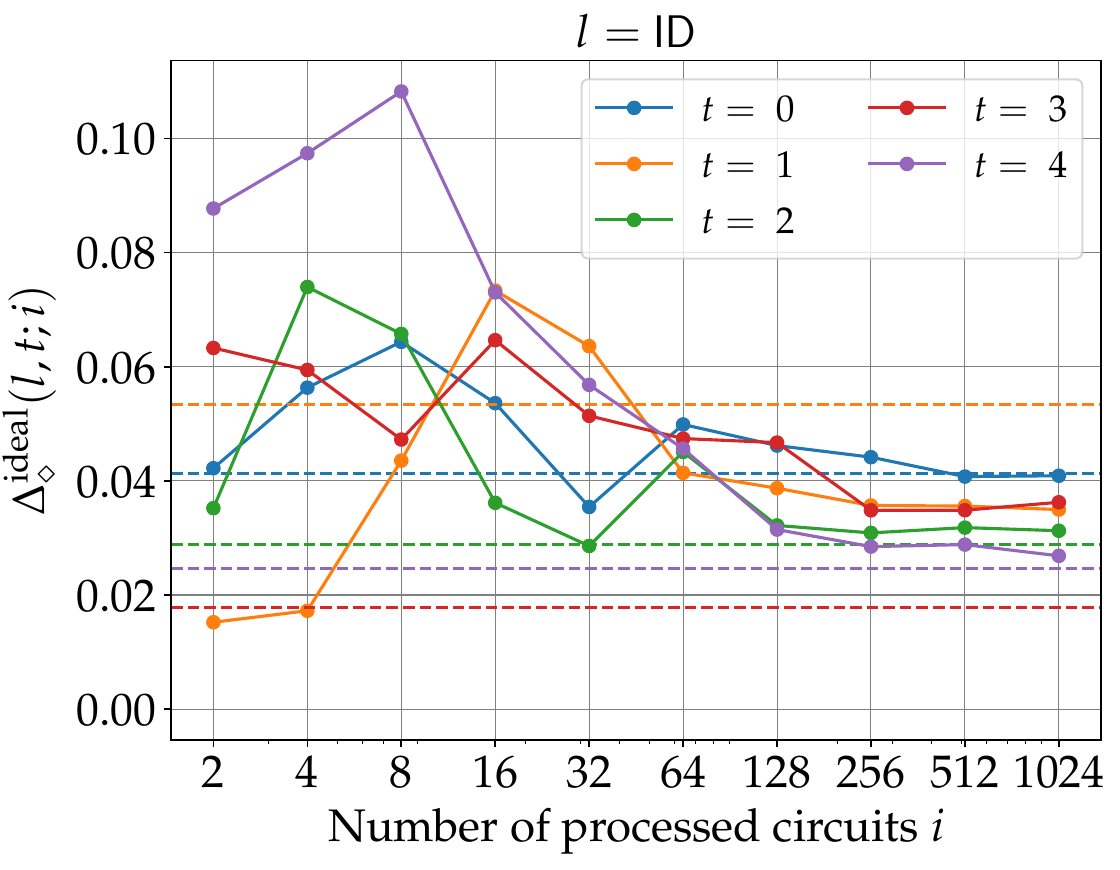}
    \includegraphics[width=0.32\linewidth]{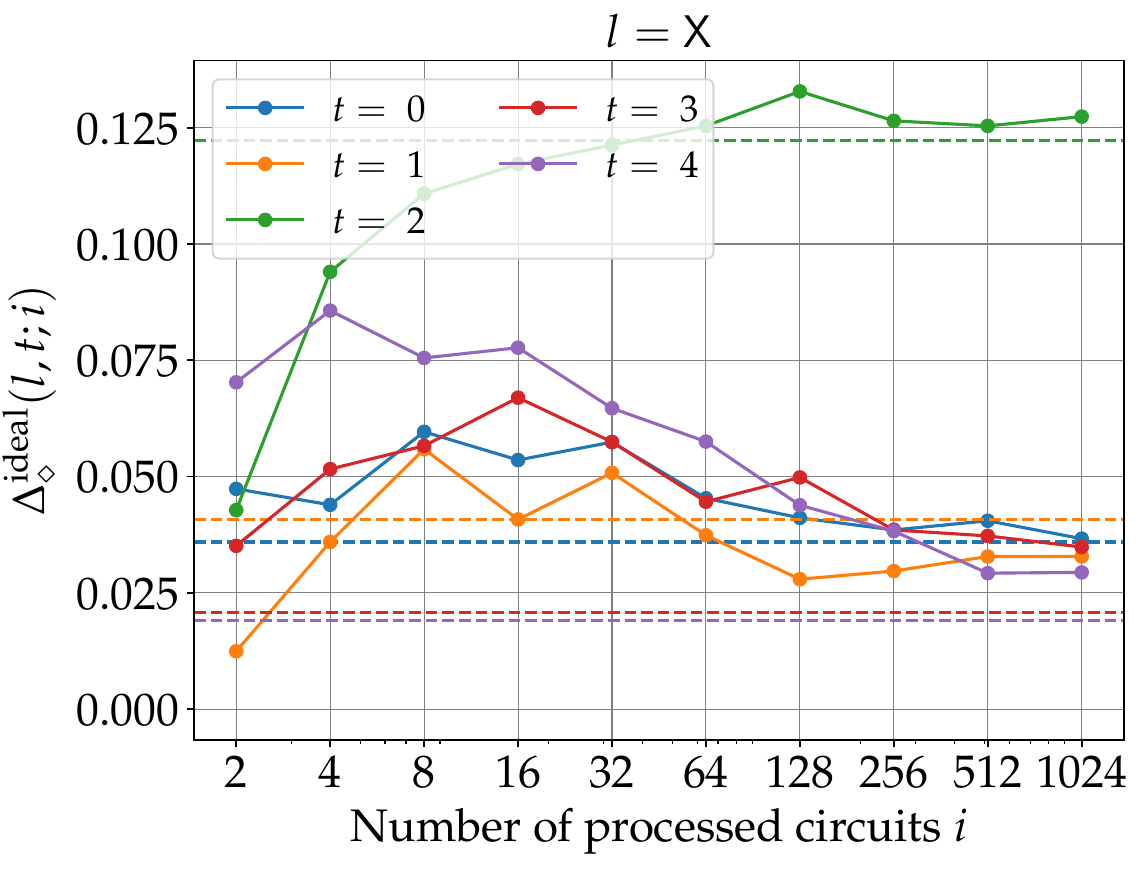}
    \includegraphics[width=0.32\linewidth]{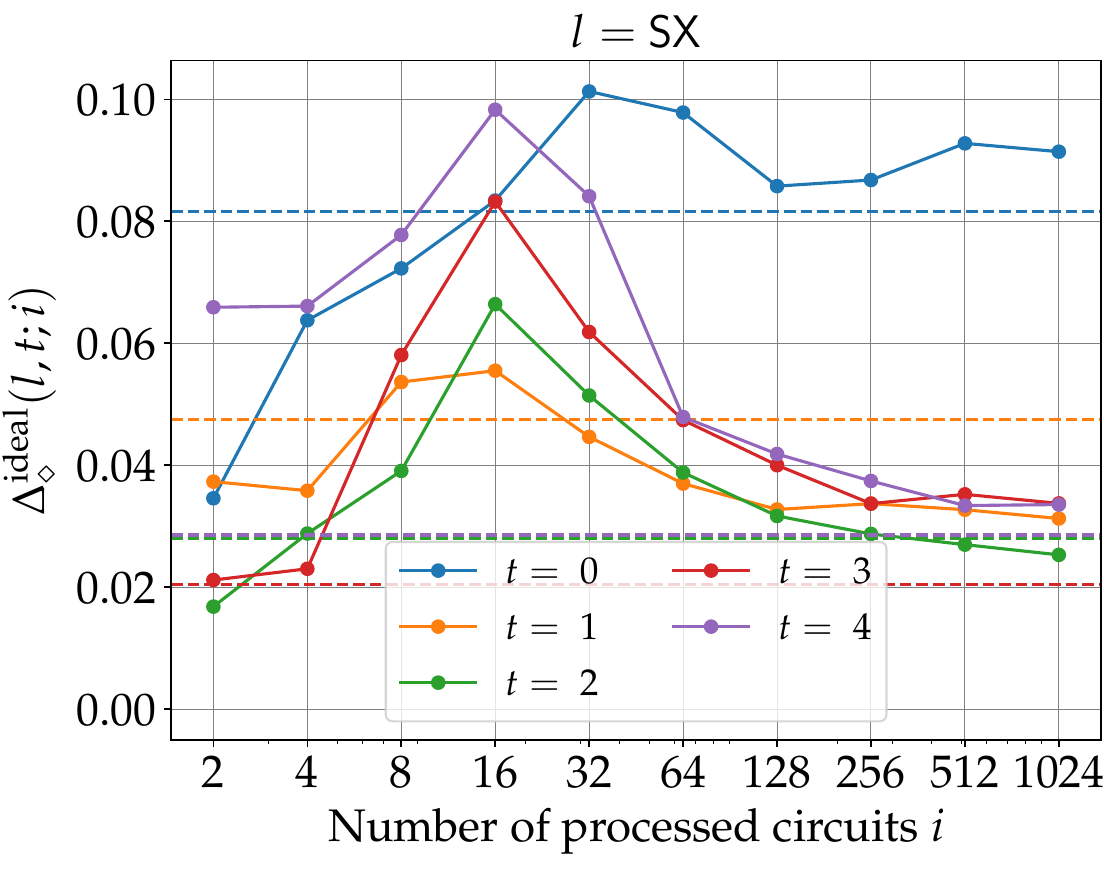}
    \includegraphics[width=0.32\linewidth]{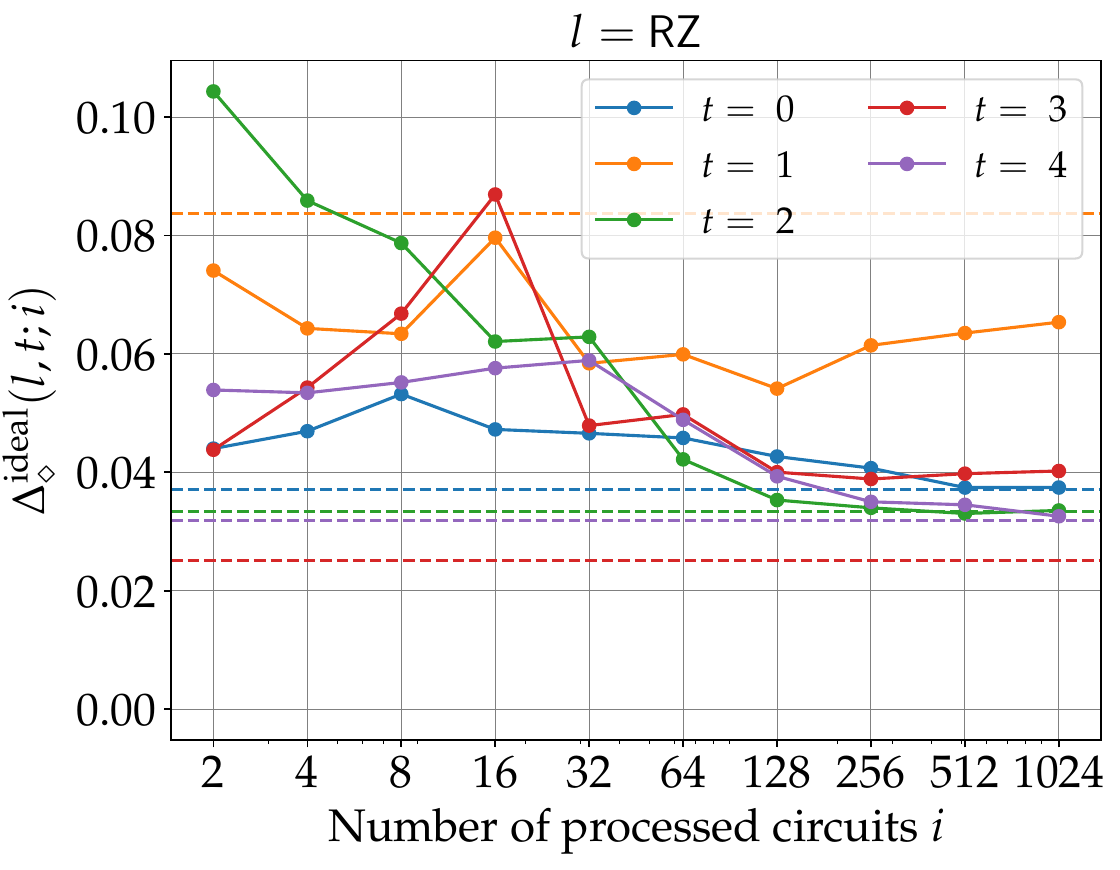}
    \includegraphics[width=0.32\linewidth]{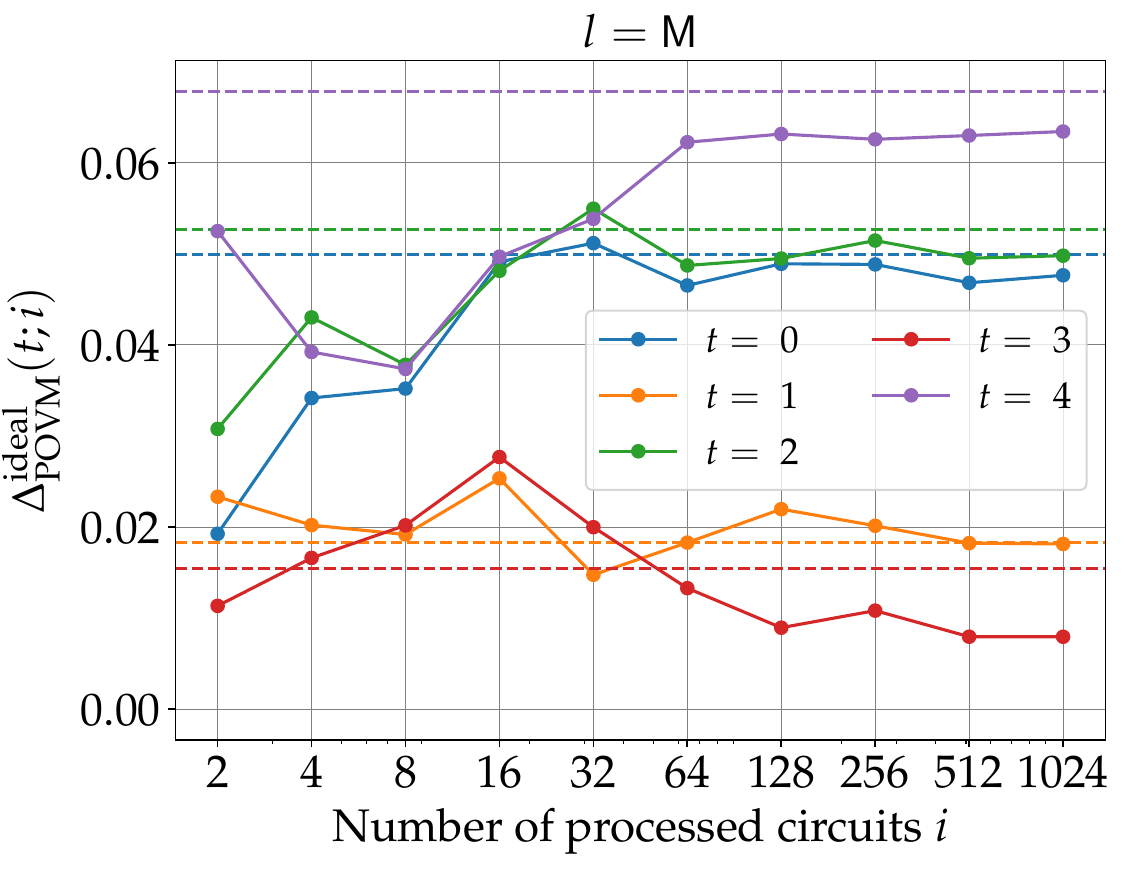}
    \includegraphics[width=0.32\linewidth]{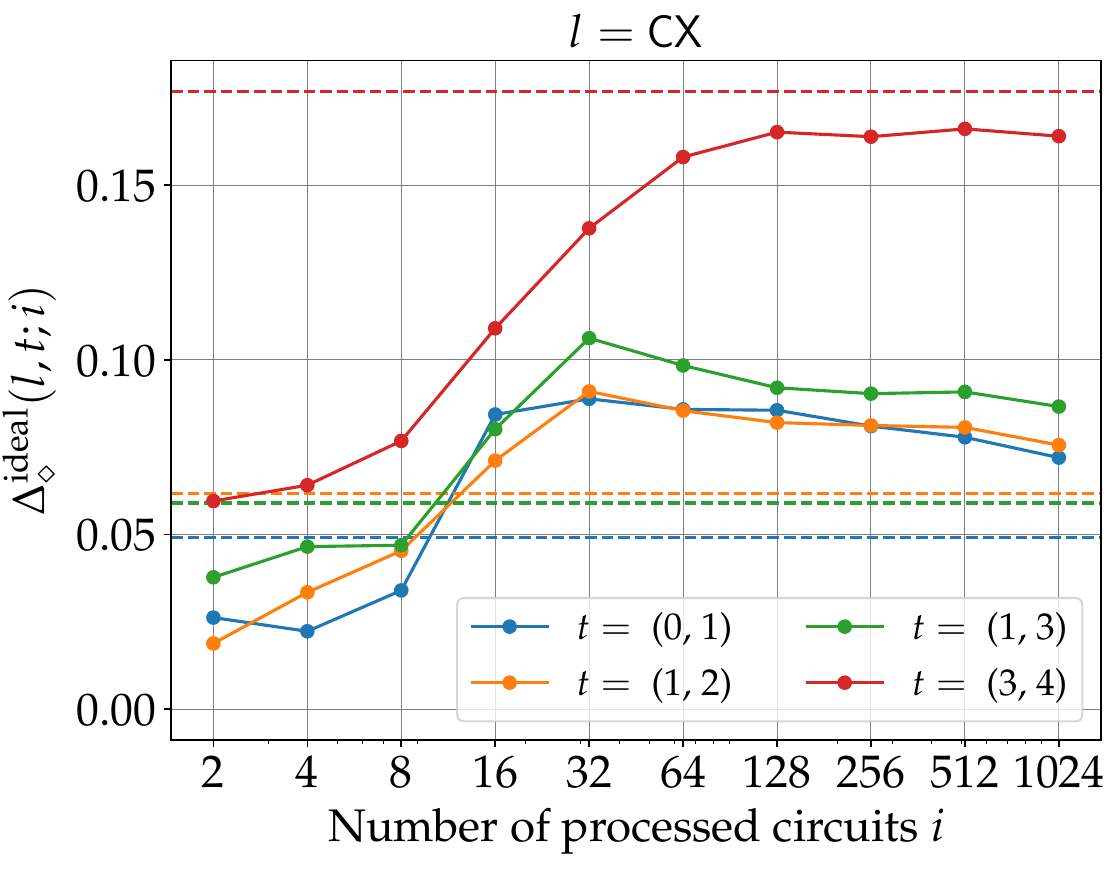}
    \caption{The distances between the ideal and estimated realizations of single-qubit gates (${\sf ID}$, ${\sf X}$, ${\sf SX}$, ${\sf RZ}$), two-qubit gates $({\sf CX}$), and read-out measurements (${\sf M}$) for different (pairs of) qubits 
    are presented as function of the number of processed circuits $i$.
    Horizontal lines show distances between the true and ideal realizations. 
    The input data for the monitoring system is generated with the described noisy quantum emulator.}
    \label{fig:synth_est_VS_ideal}
\end{figure*}

Next, we consider the distance between the estimated gate sets ${\bf G}^{\rm est}_i$ and the ideal gate set ${\bf G}^{\rm ideal}$.
The behaviour of $\Delta_\diamond^{\rm ideal}(l,t;i)$ and $\Delta_{\rm POVM}^{\rm ideal}(t;i)$ is shown in Fig.~\ref{fig:synth_est_VS_ideal}.
By the horizontal lines in Fig.~\ref{fig:synth_est_VS_ideal} we also show values of distances between the elements of the real and ideal gate sets given by
\begin{equation}
    \begin{aligned}
        &\Delta^{\rm ideal,true}(l,t)=\| {\bf G}^{\rm ideal}(l,t) - {\bf G}^{\rm true}(l,t)\|_\diamond,\\
        &\Delta_{\rm POVM}^{\rm ideal,true}(t) = \|M_0({\bf G}^{\rm ideal}({\sf M},t))-M_0({\bf G}^{\rm true}({\sf M},t)) \|_1. \\
    \end{aligned}
\end{equation}
One can see that though there are imperfections of estimates, the monitoring system allows one to identify clearly the most erroneous elements against the elements of same type 
(see e.g. ${\sf X}$ gate on 2-nd qubit, ${\sf SX}$ on 0-th qubit, and ${\sf CX}$ gate between 3-rd and 4-th qubits).

\subsection{Monitoring the real device}\label{sec:monitoring-ibmq}

Here we apply the monitoring system to analyzing samples that are obtained with the use of the IBM 5-qubit superconducting quantum processor \textsc{imbq\_lima} (see the calibration data in Appendix~\ref{sec:app:ibmq-details}).
In order to provide a fair comparison between the results obtained with the emulator and real device, we consider exactly the same input circuits, and shots number $K_i=8192$.

\begin{figure} [!ht]
    \centering
    \includegraphics[width=\linewidth]{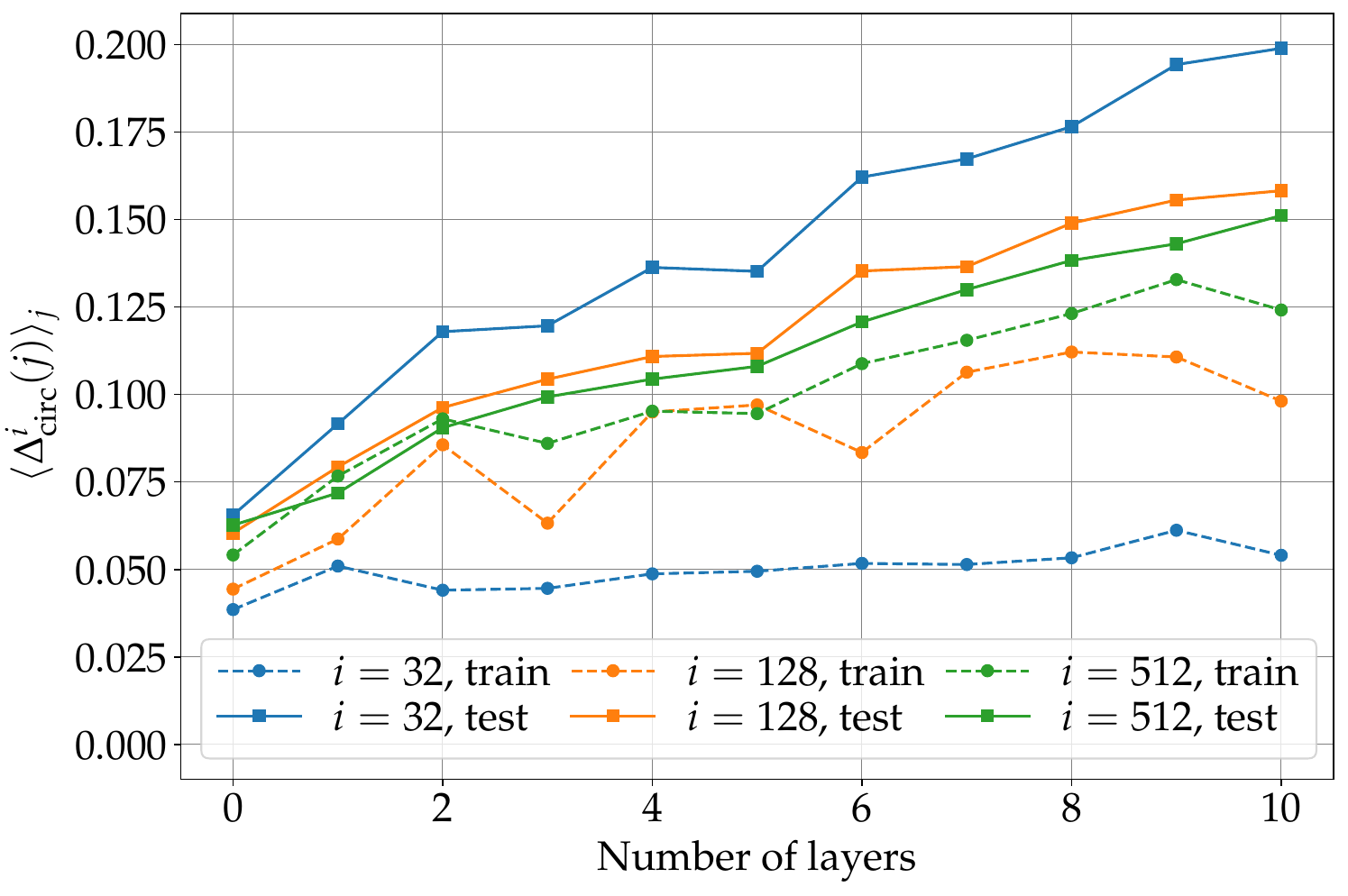}
    \caption{Mean values of prediction inaccuracy~\eqref{eq:l1_dist} obtained by averaging over circuits with the same number of layers inside the training and test sets for different number of training circuits $i$ (1024 circuits are considered in total).
    The input data for the monitoring system is obtained from \textsc{ibmq\_lima}.
    }
    \label{fig:l1norm_ibm}
\end{figure}

First, we consider an (in)accuracy of the outcomes distribution predictions as a function of a number of processed circuits.
The behaviour of averaged values of $\Delta^i_{\rm circ}(j)$, obtained in the same way as in the previous subsection, are shown in Fig.~\ref{fig:l1norm_ibm}.
We see that in general Fig.~\ref{fig:l1norm_ibm} has the same form as Fig.~\ref{fig:l1norm_synth}, yet there are some differences.
One can see in Fig.~\ref{fig:l1norm_ibm} a clear monotonic increase of the prediction inaccuracy with growth of layers number in contrast to a saturation behaviour observed for the synthetic data.
At the same time, the values of $\Delta^i_{\rm circ}(j)$ for the data from the real processor are 2-3 times higher than for the synthetic data.
We suppose that this increase of discrepancies is due to the fact the the suggested model of the quantum processor does not fully catch all the features of real processor.
However, as we show next, the monitoring system still provides valuable results.

\begin{figure*} [!ht]
    \centering
    \includegraphics[width=0.32\linewidth]{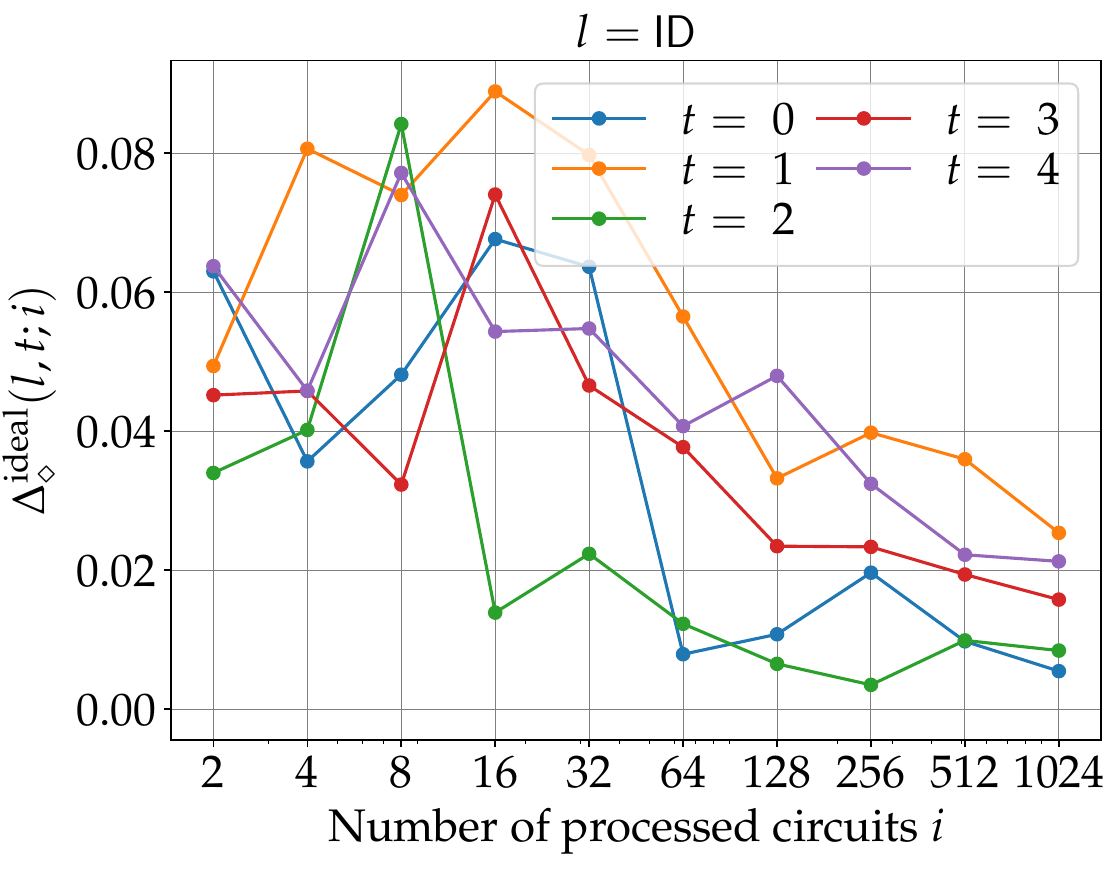}
    \includegraphics[width=0.32\linewidth]{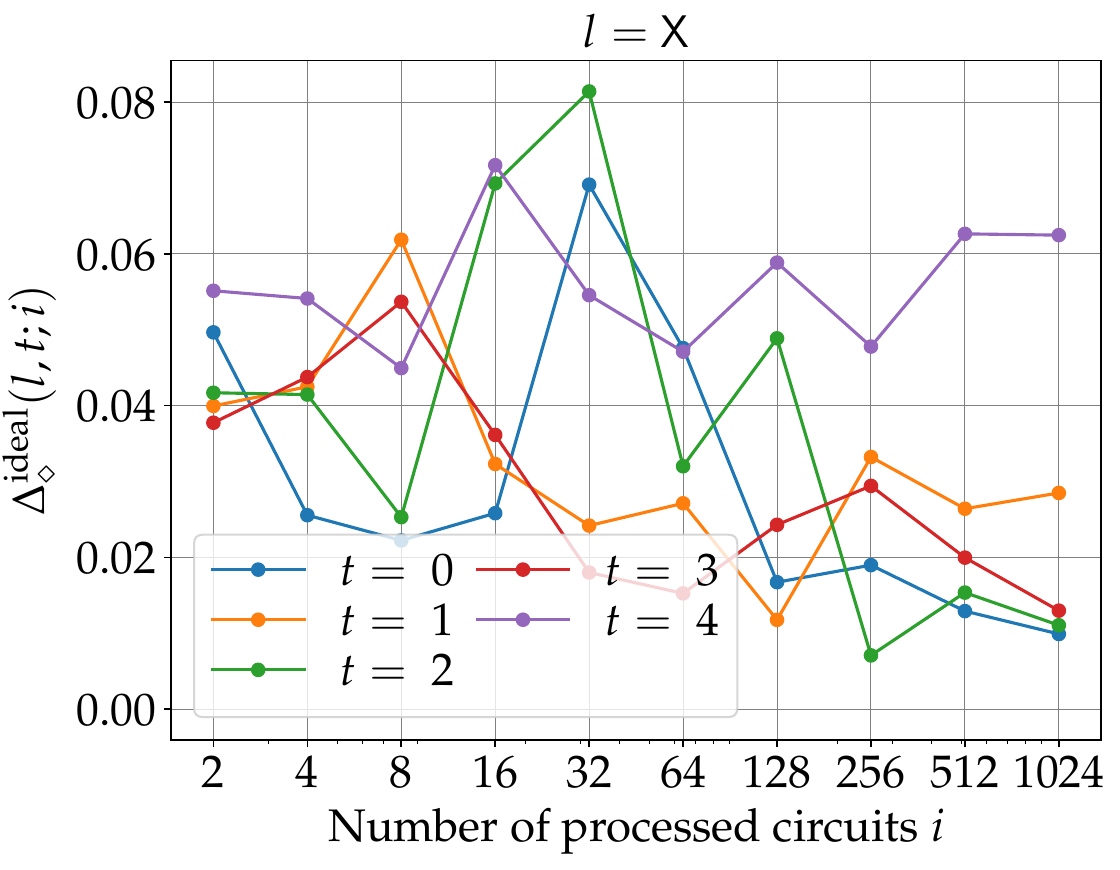}
    \includegraphics[width=0.32\linewidth]{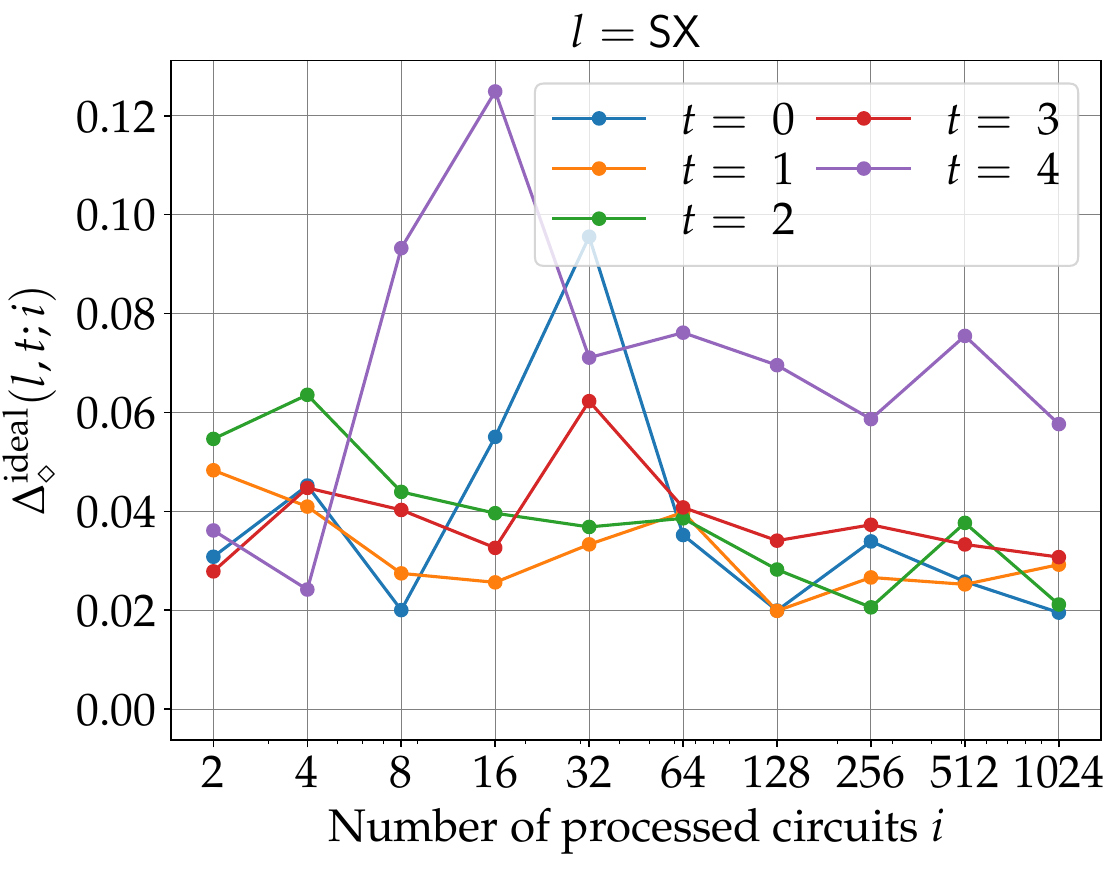}
    \includegraphics[width=0.32\linewidth]{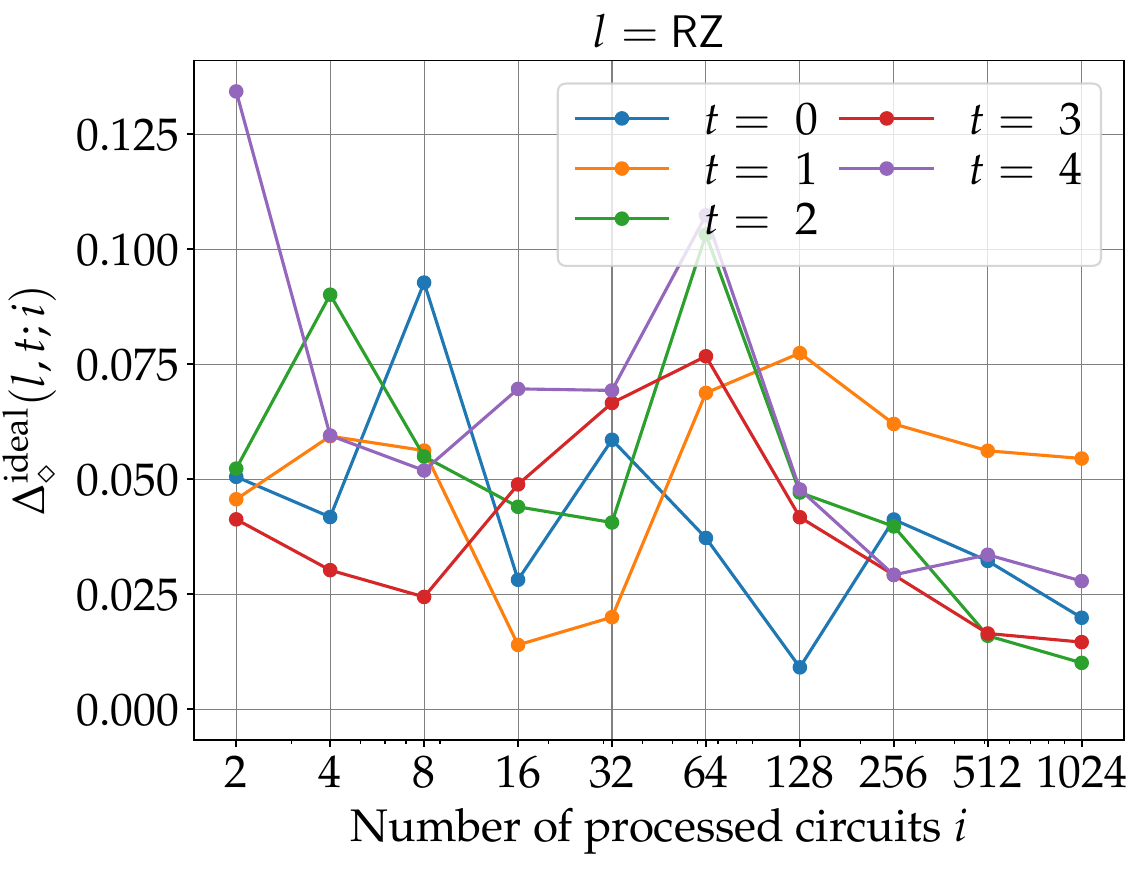}
    \includegraphics[width=0.32\linewidth]{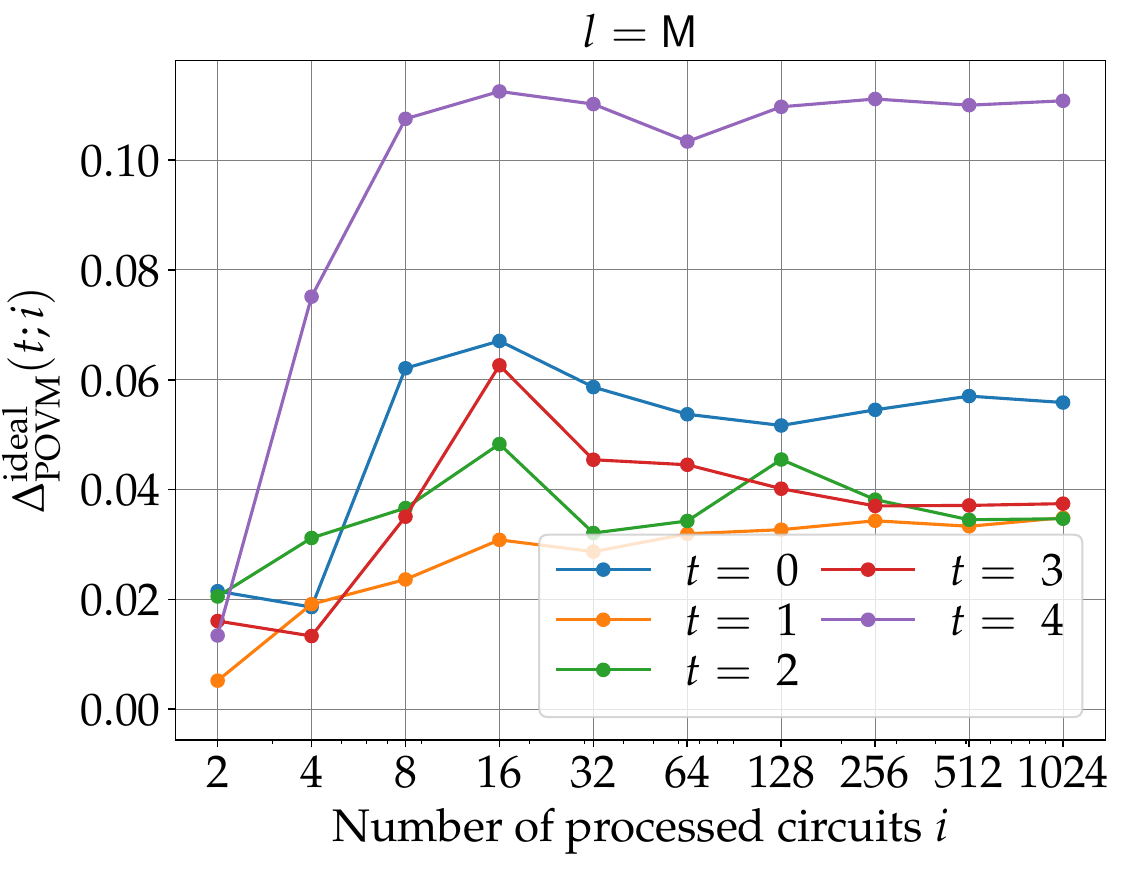}
    \includegraphics[width=0.32\linewidth]{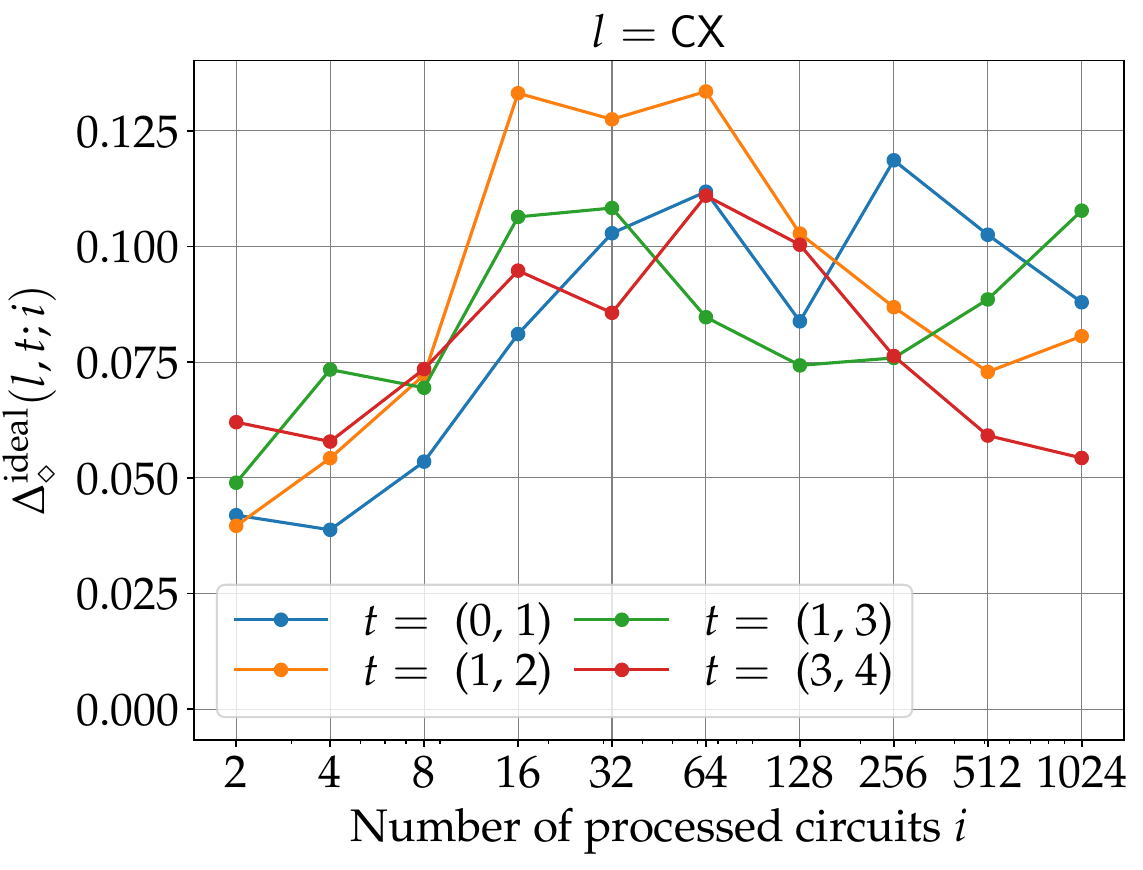}
    \caption{
    The distances between the ideal and estimated realizations of single-qubit gates (${\sf ID}$, ${\sf X}$, ${\sf SX}$, ${\sf RZ}$), two-qubit gates $({\sf CX}$), and read-out measurements (${\sf M}$) for different (pairs of) qubits are shown
    as function of the number of processed circuits $i$.
    % Horizontal lines show distances between the true and ideal realizations. 
    The input data for the monitoring system is obtained from \textsc{ibmq\_lima}.}
    \label{fig:ibm_est_VS_ideal}
\end{figure*}

In Fig.~\ref{fig:ibm_est_VS_ideal} we show distances between elements of the estimated gate sets ${\bf G}^{\rm est}_i$ and corresponding ideal elements from ${\bf G}^{\rm ideal}$.
One can see that the realization of ${\sf M}$, ${\sf X}$, and ${\sf SX}$ gates on 4-th qubit ($t=4$) clearly stands out from the corresponding realizations of other qubits.
We can conclude that it is the 4-th qubit that requires a calibration first.

In order to validate our conclusions about the quality of operation on 4-th qubit, we launch an additional standard read-out measurement calibrations routine.
It consists of preparing and measuring each of 32 5-qubit computational basis states ($\ket{00000}, \ldots, \ket{11111}$) given number of times (in our realization it equals to 8192).
In this way we run 32 additional circuits with 8192 shot each.
In the result we obtain a $32\times 32$ `calibration matrix' whose entries are frequencies of obtaining a particular outcome for particular input state.
Using the calibration matrix, we extract experimental error rates $e_{0\rightarrow 1}^{\rm exp}(t)$ and $e_{1\rightarrow 0}^{\rm exp}(t)$, where 
$e_{x\rightarrow y}^{\rm exp}(t)$ is a probability (frequency) of measuring the result $y$ for the prepared state $x$ on $t$-th qubit.

At the same time we take the result of the monitoring system after processing 1024 circuits ${\bf G}^{\rm est}_{1024}$ and estimate the experimental rate in the form
\begin{equation}
    e_{x\rightarrow y}^{\rm est}(t) = \bra{x}M_x\left({\bf G}^{\rm est}_{1024}({\sf M},t)\right)\ket{y}.
\end{equation}
We can make the estimate of $e_{1\rightarrow 0}(t)$ more accurate by taking into account that state $\ket{1}$ is obtained from $\ket{0}$ by applying ${\sf X}$ gate.
The more accurate estimate then takes the form
\begin{equation}
    e_{1\rightarrow 0}^{\rm est, acc}(t) = {\rm Tr}\left[{\bf G}^{\rm est}_{1024}({\sf X},t)[\ket{0}\bra{0}]
    M_0\left({\bf G}^{\rm est}_{1024}({\sf M},t)\right)\right].
\end{equation}

\begin{figure} [!ht]
    \centering
    \includegraphics[width=0.9\linewidth]{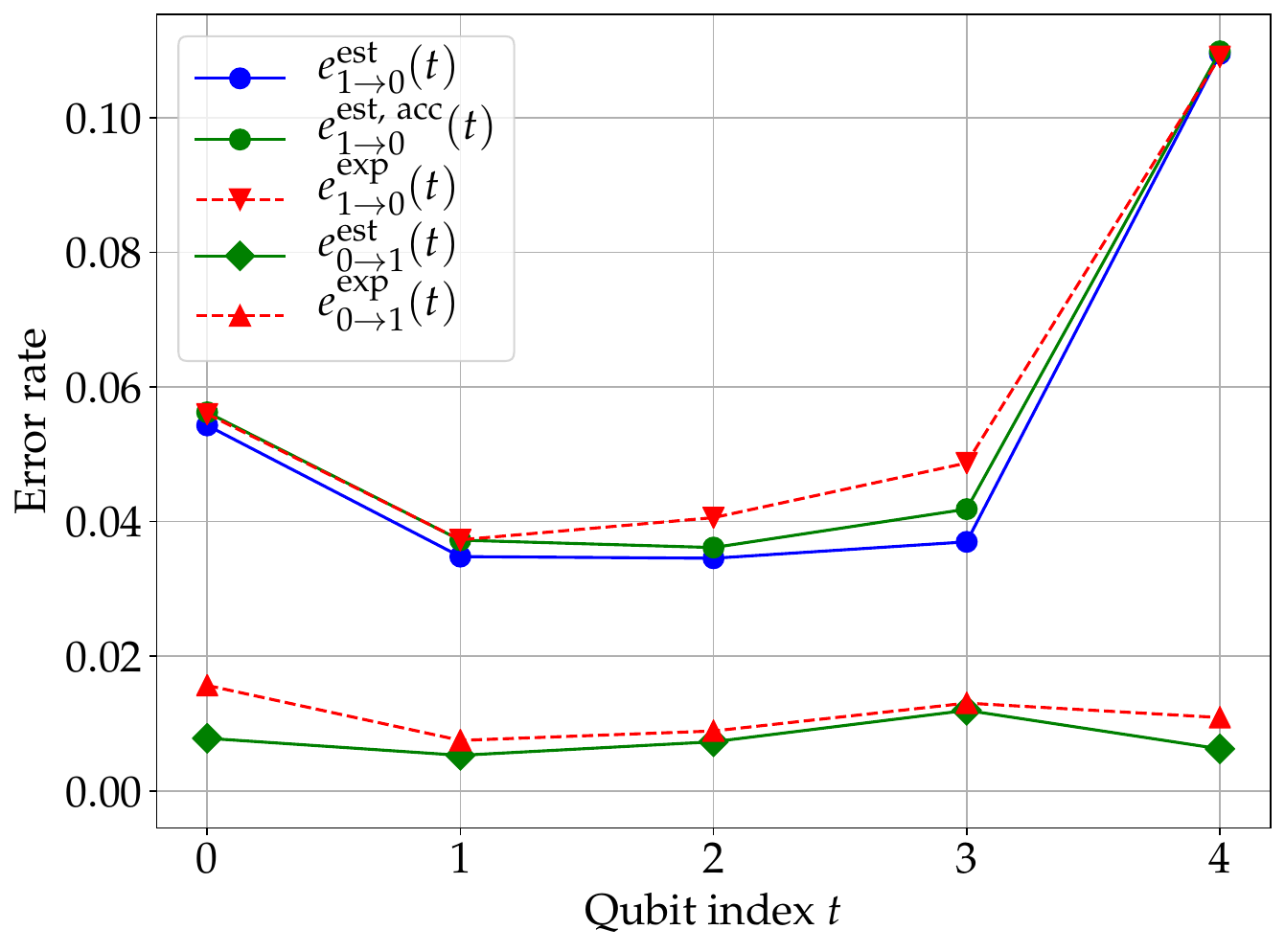}
    \caption{Comparison of read-out measurement error rates obtained from the specially launched calibration routine [$e^{\rm exp}_{1\rightarrow 0}(t)$ and $e^{\rm exp}_{0\rightarrow 1}(t)$] 
    and the monitoring system trained on 1024 random circuits [$e^{\rm est}_{1\rightarrow 0}(t)$, $e^{\rm est, acc}_{1\rightarrow 0}(t)$, and $e^{\rm est}_{0\rightarrow 1}(t)$] is presented.}
    \label{fig:mon_vs_cal}
\end{figure}

The comparison between the results extracted from the calibration matrix and the monitoring system is presented in Fig.~\ref{fig:mon_vs_cal}.
We see a good match between them, and, as we expect, $e_{1\rightarrow 0}^{\rm est, acc}(t)$ is closer to $e_{1\rightarrow 0}^{\rm exp}(t)$ than $e_{1\rightarrow 0}^{\rm est}(t)$.
Thus the monitoring system trained on results from random circuits provides almost the same results as the specially launched calibration routine without borrowing any additional computational time.

\section{Conclusion and outlook} \label{sec:concl}

In the present work, we have developed the monitoring system that allows one to get adequate estimates of basic elements of a quantum processor based on samples obtained after running random circuits.
The main feature of the developed system in comparison with other benchmarking techniques, i.e. randomized benchmarking and gate set tomography, 
is that it does not borrow any additional computational time from the quantum processor and can be applied to arbitrary circuits that are not under control (e.g., in the case where input circuits come from remote users).
The source code of the developed system is available on~\cite{monitoring_code}.

We have tested the developed system using a synthetic data generated from a specifically designed emulator of a noisy quantum processor, and data obtained by launching circuits on the 5-qubit superconducting quantum processor \textsc{ibmq\_lima}, 
which is available via cloud.
In the case of the synthetic data, we have validated the performance of the monitoring system by comparing the output of the system (estimated gate set) with the true noisy gate set realized within the emulator.
For the case of the data obtained from the quantum processor, we have seen the results of the monitoring system are in a pretty good agreement with the results obtained from specially launched read-out calibration routine.
Thus we can conclude that the developed system successfully copes with assigned tasks.

The main limitation of the applicability of the considered approach relates to the necessity of inherent quantum emulator, which in fact mimics the behaviour of real quantum processor. 
This issue prevents the use of the monitoring system for processing large circuits intractable for tensor network-based simulation.
However, the monitoring system is still applicable in the case where the set of launched circuits possess a subset of tractable, i.e. of low width or low depth, circuits.
A prime example of this kind of circuits are `elided' and `patch' circuits with bounded amount of generated entanglement used for verification purposes in quantum advantage experiments~\cite{Martinis2019}.
In this way, even in a regime, where emulation time is larger then the real quantum computing time, the developed system still is able to provide valuable information by processing subsets of launched circuit.
This subset can chosen according to expected emulation complexity (e.g., circuit width $\times$ circuit depth), or taken at random (e.g., by processing every $k$th launched circuit for large enough $k$).

%In this case, the processing of the subset of launched tractable circuits can provide valuable information about elements of the quantum processor.

Another limitations of the developed system relate to inherent assumptions about the noise model.
One of the used assumptions is Markovianity of noise in the sense that each noisy gate can be described with a CPTP map, 
and we also assume that noisiness of a gate acting on a particular qubit (pair of qubits) does not depend on the position if this gate within a circuit.
We note that the first limitation can be overcome by introducing additional quantum information carries (i.e. qubits or qudits) that are responsible for memory effects.
Then gates from input circuits can be mapped to quantum channels acting both on the original target and additional quantum information carries.
This approach relates to the description of non-Markovian behaviour by using a Markovian embedding technique~\cite{budini2013embedding,campbell2018system,luchnikov2020machine}.
We leave a support of non-Markovian processes for further development of the monitoring system.

Finally, we note the developed system is capable of tracking drifts of noise parameters in the case where these drift appears on time scales larger than the number of circuits reliable training (based on the obtained results it is about $2^6 \ldots 2^8$ circuits). 
This tracking is possible by adjusting the hyperparameter $M$, defining the maximal number of recently launched circuits taken for input for the monitoring system.
We expect that our monitoring system can become a useful tool for different quantum computing platforms, reducing additional resources needed for their calibration. 

\section*{Acknowledgments}
We acknowledge use of the IBM Q Experience for this work. The views expressed are those of the authors and do not reflect the official policy or position of IBM or the IBM Q Experience team. 
The work of I.A.L., J.A.L., and E.O.K. (the concept of the monitoring system, analysis of the results) was supported by the Russian Science Foundation Grant No. 19-71-10091. 
The work of Y.F.Z. and A.K.F. was supported by the Russian Roadmap on Quantum Computing (tests of the monitoring system; Contract No. 868-1.3-15/15-2021, October 5, 2021).
We also thank the Priority 2030 program at the National University of Science and Technology ``MISIS” under the project K1-2022-027 (benchmarking quantum emulator).

\appendix

\section{Samples generation} \label{sec:app:emulator}

The operation of the quantum processor emulator used for generating synthetic data requires an efficient algorithm of measurement outcomes sampling. 
Remind that each measurement outcome is a bit string $s = (s_0, s_1, \dots, s_{n-1})$, where $s_i$ takes either $1$ or $0$.
The probability of getting a particular measurement outcome reads
\begin{equation}
    P(s_{n-1}, s_{n-2}, \dots, s_0) = {\rm Tr}\left(\bigotimes_{i=0}^{n-1}\ket{s_i}\bra{s_i}\rho_{\rm f}\right),
\end{equation}
where $\rho_{\rm f}$ is the final state of a quantum processor after a sequence of instructions and $\ket{s_i}\bra{s_i}$ is a projector acting on the $i$-th qubit. Straightforward sampling from $P(s_{n-1}, s_{n-2}, \dots, s_0)$ requires an appeal to all possible bit strings whose number scales exponentially with $n$. 
To circumvent such scaling we utilize chain rule for probabilities that reads
\begin{equation}
    \label{eq:chain_rule}
    P(s_{n-1}, s_{n-2}, \dots, s_0) = \prod_{i=0}^{n-1} P(s_i|s_{i-1}, \dots, s_0),
\end{equation}
where conditional probabilities take the following form:
\begin{equation}
    \begin{aligned}
    P(s_i|s_{i-1}, \dots, s_0) &= \frac{P(s_i,s_{i-1}, \dots, s_0)}{\sum_{x=0,1}P(x,s_{i-1}, \dots, s_0)},\\
    P(s_i,s_{i-1}, \dots, s_0) &= {\rm Tr}\left(\bigotimes_{j=i+1}^{n-1} \mathbb{1} \bigotimes_{j=0}^{i} \ket{s_i}\bra{s_i} \rho_{\rm f} \right),
    \end{aligned}
\end{equation}
where $\mathbb{1}$ is the identity matrix reflecting the fact that we have not yet observed the measurement outcomes for qubits number $\{i+1, i+2, \dots, n-1\}$. 

The decomposition Eq.~\eqref{eq:chain_rule} allows one to sample measurement outcomes sequentially, i.e. at the first step one samples a measurement outcome for the first qubit from the distribution $P(s_0)$, 
at the second step on samples a measurement outcome for the second qubit given the measurement outcome for the first qubit from the distribution $P(s_1|s_0)$ and so on. 
The complexity of this sequential strategy scales linearly with $n$ in contrast with the exponential scaling for the straightforward approach.

\section{Quantum emulator benchmarks} \label{sec:app:benchs}

Here we provide an additional information about the performance of the tensor network-based quantum emulator employed in the developed monitoring system.
We benchmark the emulator on $n$-qubit random circuits consisted of noisy gates, described by corresponding quantum channels (CPTP maps).
Circuits structured in layers are considered.
Each layers consists of single qubit gates applied to each of $n$ qubits, and two-qubit gates applied to $\lfloor n/2 \rfloor$ randomly chosen disjoint qubit pairs.
The considered native gates set consists of ${\sf ID}$, ${\sf X}$, ${\sf SX}$, ${\sf RZ}$, ${\sf M}$, and ${\sf CX}$ gates.
All-to-all connection topology for ${\sf CX}$ gates is assumed.

\begin{figure}
    \centering
    \includegraphics[width=\linewidth]{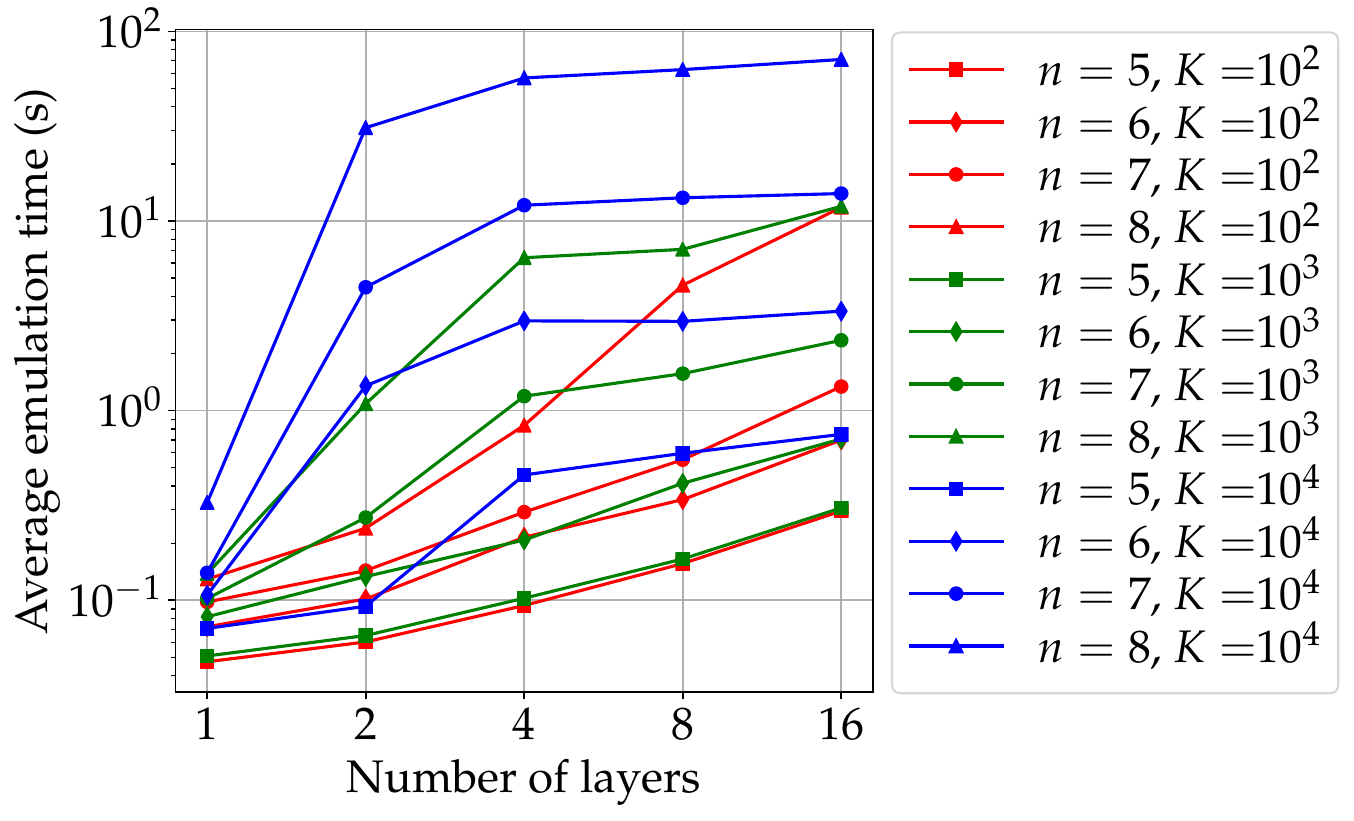}
    \caption{Average emulation time of noisy random circuits using the employed tensor network-based emulator.
    The averaging is made over 5 independent numerical experiments.}
    \label{fig:bench}
\end{figure}

An average time for obtaining $K$ outcomes as function of $n$, $K$, and a number of layers is shown in Fig.~\ref{fig:bench}.
AMD Ryzen 5 3600 with 16 Gbyte RAM is used.
We note that no optimizations related to GPU-based parallelization are present in the current version of the emulator.
We also recall that the emulation is performed without any tensor rank contractions usually considered in the framework of tensor network-based representations.

\section{Reconstructing single-qubit gates from random circuits}

Here we provide an additional numerical experiment of reconstructing noisy quantum gates with the developed monitoring system.
In particular, we consider a single-qubit processor with a universal gate set consisting of ${\sf RX}$ and ${\sf RZ}$ gates providing $\pi/4$ rotations around $x$- and $z$-axes of the Bloch sphere correspondingly.
The input data for the monitoring system is obtained by running random circuits on the developed noisy emulator, described in Sec.~\ref{sec:emulator}.
The chosen noise whose parameters are listed in Table~\ref{tab:noiseforappendix}.

\begin{figure}
    \centering
    \includegraphics[width=\linewidth]{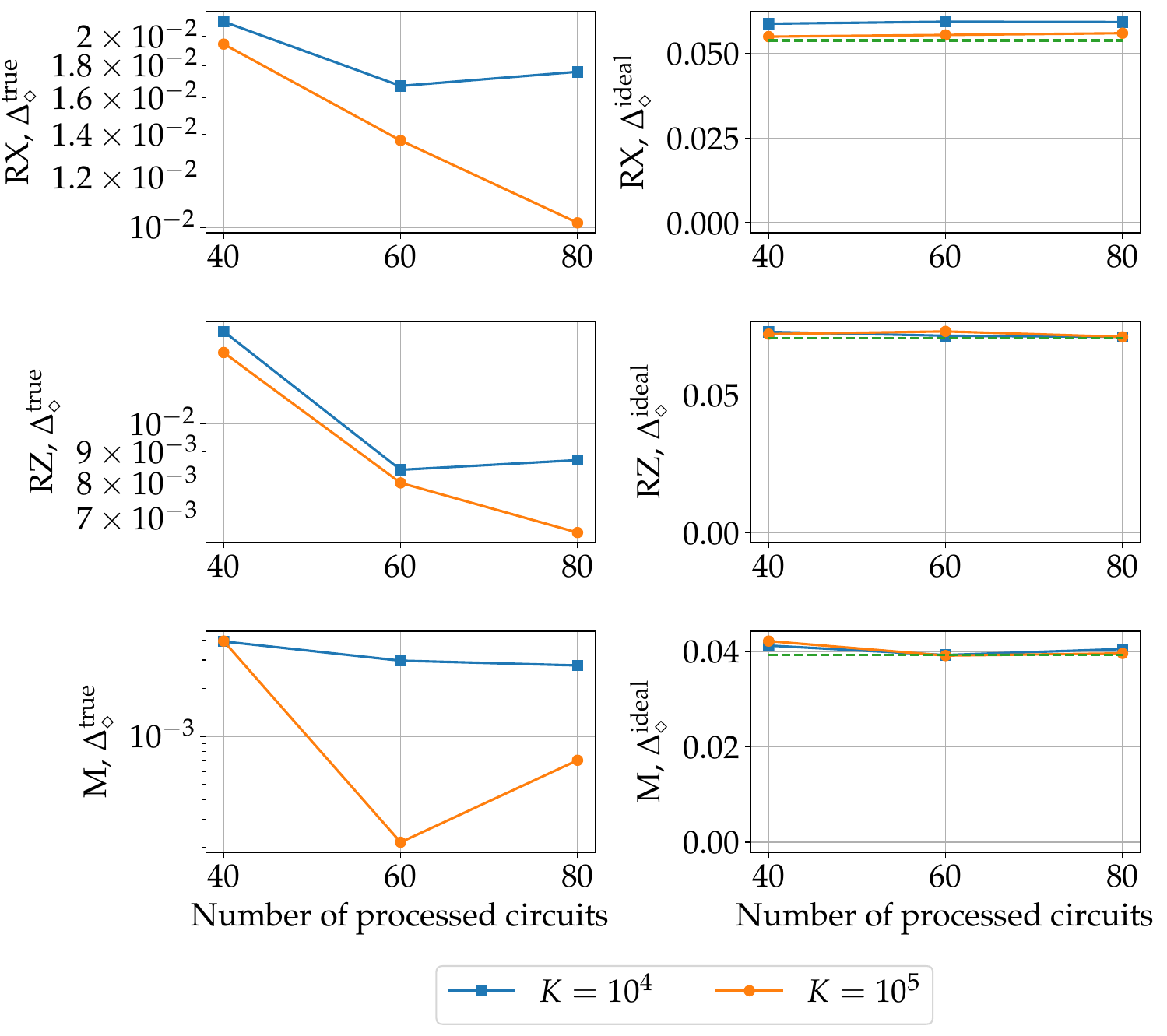}
    \caption{
    The distances between the true and the estimated (left column), and the ideal and the estimated (right column) realizations of single-qubit gates ($\sf RX$, $\sf RZ$), and read-out measurements (${\sf M}$).
    Horizontal lines in the right column show distances between the true and ideal realizations.}
    \label{fig:singlequbit_case}
\end{figure}

The results of noisy channels reconstruction based on running random circuits are shown Fig.~\ref{fig:singlequbit_case}.
Recall that the reconstruction in the monitoring system is performed with respect to general quantum channels completely independent of the noise model in the emulator used for data generation.
The number of gates for each circuit is taken at random from the range $\{1,2,\ldots,30\}$, and each gate within a circuit is taken at random from the set $\{{\sf RX}, {\sf RZ}\}$.
The number of runs of each circuit $K=10^4$ and $K=10^5$ are considered.

One can clearly see that the reconstruction accuracy increases with a growth of $K$ and a number of processed circuits.
Notably, the obtained accuracy values are better than the ones shown in Fig.~\ref{fig:synth_est_VS_real}.
This behaviour can be explained by reduced number of reconstructed noisy gates and absence of two-qubit gates.

\begin{table}[htp]
    \centering
    \begin{tabular}{c|c|c|c|c|c}
         $l$ & $\nu^l$ & $p^l$ &  $\mu^l$ & $\gamma^l$ & $F^l$  \\ \hline \hline
         ${\sf RX}$ & 0.065 & 0.03 & 0.005 & 0.005 & 0.971 \\ \hline
         %${\sf S}$ & 0.02 & 0.02 & 0.02 & 0.02 & 0.970 \\ \hline
         ${\sf RZ}$ & 0.06 & 0.01 & 0.03 & 0.03 &  0.970 \\
         \hline
         ${\sf M}$ & 0 & 0.04 & 0.02 & 0.02 & 0.956 \\
    \end{tabular}
    \caption{Noise parameters of the emulator and resulting fidelities for the considered single-qubit case.}
    \label{tab:noiseforappendix}
\end{table}

\section{Calibration data} \label{sec:app:ibmq-details}

In Table~\ref{tab:calibration} we provide additional calibration data for the employed  \textsc{ibmq\_lima} processor at the time of the demonstration.
This demonstration data was downloaded using cloud-based interface.

\begin{table*}[htp]
    \centering
    \begin{tabular}{c|c|c|c|c|c}
    Parameter & {\sf Q}0 & {\sf Q}1 & {\sf Q}2 & {\sf Q}3 & {\sf Q}4 \\\hline \hline
    $T_1$ ($\mu$s) & 117.49 & 66.93 & 121.65 & 93.56 & 21.35 \\ \hline
    $T_2$ ($\mu$s) & 184.91 & 79.02 & 182.49 & 93.73 & 25.86 \\ \hline
    Frequency (GHz) & 5.029 & 5.127 & 5.247 & 5.302 & 5.092 \\ \hline 
    Anharmonicity (GHz) & -0.33574 & -0.31834 & -0.3336 & -0.33124 & -0.3429 \\\hline
    Readout assignment error & 0.0297 & 0.0133 & 0.0203 & 0.0267 & 0.0484 \\ \hline
    Prob. meas 0 prep. $\ket{1}$ & 0.0466 & 0.0216 & 0.0328 & 0.0374 & 0.0752 \\ \hline
    Prob. meas 1 prep. $\ket{0}$ & 0.0128 & 0.005 & 0.0078 & 0.016 & 0.0216 \\ \hline
    Readout length (ns) & 5351.11 & 5351.11 & 5351.11 & 5351.11 & 5351.11 \\ \hline
    ${\sf ID}$ error  & $2.039~10^{-4}$  & $2.333~10^{-4}$  & $2.414~10^{-4}$  & $2.759~10^{-4}$  & $6.108~10^{-4}$  \\ \hline
    ${\sf SX}$ error  &	$2.039~10^{-4}$ &	$2.333~10^{-4}$ &	$2.414~10^{-4}$ &	$2.759~10^{-4}$ &	$6.108~10^{-4}$ \\ \hline
    ${\sf X}$ error  &	$2.039~10^{-4}$ &	$2.333~10^{-4}$ &	$2.414~10^{-4}$ &	$2.759~10^{-4}$ &	$6.108~10^{-4}$ \\ \hline
    \end{tabular}\\
    \vspace{5pt}{\sf CX} error\\
    \begin{tabular}{c|c|c|c|c|c}
        & {\sf Q}0 & {\sf Q}1 & {\sf Q}2 & {\sf Q}3 & {\sf Q}4\\ \hline
         {\sf Q}0 & & $5.927~10^{-3}$ & & & \\ \hline
         {\sf Q}1 & $5.927~10^{-3}$ & & $5.744~10^{-3}$ & $1.182~10^{-2}$ & \\ \hline
         {\sf Q}2 & & $5.744~10^{-3}$ & &\\ \hline
         {\sf Q}3 & & $1.182~10^{-2}$ & & & $1.493~10^{-2}$ \\ \hline
         {\sf Q}4 & & & & $1.493~10^{-2}$ &  \\
    \end{tabular}\\
    \vspace{5pt}Gate time (ns)\\
    \begin{tabular}{c|c|c|c|c|c|}
        & {\sf Q}0 & {\sf Q}1 & {\sf Q}2 & {\sf Q}3 & {\sf Q}4\\ \hline
         {\sf Q}0 & & 305.777 & & & \\ \hline
         {\sf Q}1 & 341.333 & & 334.222 & & \\ \hline
         {\sf Q}2 & & 298.666  & & & \\ \hline
         {\sf Q}3 & & 462.222 & & & 519.111   \\ \hline
         {\sf Q}4 & & & & 483.555 &  \\ 
    \end{tabular}
    \caption{Calibration data for the five-qubit superconducting processor \textsc{ibmq\_lima} at the time of the demonstration discussed in Sec.~\ref{sec:monitoring-ibmq}.}
    \label{tab:calibration}
\end{table*}

%\clearpage

\bibliography{bibliography.bib}

\end{document}